\documentclass[nofootinbib,onecolumn]{revtex4-2}
\usepackage{relsize}
\usepackage{upgreek}
\usepackage{placeins}
\usepackage{amsmath,amssymb,accents}%,graphicx,hyperref}
\usepackage{titlesec}
\usepackage[margin=1in]{geometry}
\usepackage{mathtools}
\usepackage{bm}% bold math5
\newcommand{\pd}{\partial}
\def\d{\mathrm{d}}
\def\Re{\mathrm{Re}}
\def\Im{\mathrm{Im}}

\newcommand{\cO}{\mathcal{O}}
\def\acosh{\text{acosh}}

\begin{document}
\title{Images of point charges in conducting ellipses and prolate spheroids}
\author{Matt Majic} \email{mattmajic@gmail.com}
\affiliation{The MacDiarmid Institute for Advanced Materials and Nanotechnology,
	School of Chemical and Physical Sciences, Victoria University of Wellington,
	PO Box 600, Wellington 6140, New Zealand}
\date{\today}

\begin{abstract}
This is an investigation into the exact forms of images of point charges in 2D conducting ellipses and 3D prolate spheroids.
For an ellipse with exterior point source, we compare two previous expressions for the analytic continuation inside the ellipse down to the focal line, the location of the image charge. For an interior source, we discover a sequence of image charges lying outside.
For a point source close the surface of a spheroid, series solutions for the potential diverge in a region that encompasses the singularities of the continuation of the reflected potential inside the spheroid.
We uncover an image system for a point charge on the axis of a prolate spheroid that extends from the focal segment somewhat further up the axis. For an off axis charge we find a new approximate image charge.  For a point charge at the center of the spheroid, the image singularity is found to lie on an infinitely wide, flat sheet with a hole around the spheroid. For a point charge located anywhere inside the spheroid, we find two new approximate image charge solutions, a point charge in real space which applies when source lies near the surface, or a point charge in complex space which projects as a ring in the real space which applies when the source lies near the rotation axis. When the point charge lies exactly on a focal point, a multipole of order 1/2 is an ideal image approximation.
\end{abstract}
\maketitle

\section{Introduction}

The image solution for the electrostatic point charge near a conducting sphere is well known and consists of an image point charge located at the Kelvin inversion point. It is natural to ask how this might extend to similar analytic geometries such as the spheroid. In other words, our goal is to analytically continue Green's function inside the spheroid as far as we can -- representing the function analytically everywhere except on the smallest possible singularity, and this singularity we call the ``image" of the point charge.
This is equivalent to a Dirichlet problem where an analytic function must be found that has the $1/r$ singularity in the vicinity of the point charge and is zero on the boundary of the spheroid.
%An approximate successive image technique was proposed in \cite{chow1979static} 
%The first paper to study this seems to be from as recent as 1984 \cite{Redzic1994}, but used the incorrect assumption of a point image. 
The first exact formulation of the image of a point charge on the axis of a spheroid was in 1995 - partial success was found by Sten and Lindell \cite{sten1995electrostatic} where the image consisted of a line charge lying on the focal segment. A later investigation of the  image solution for the elliptic cylinder \cite{sten1996focal}, revealed that the image lies on a focal strip, but if the source is close enough, a disjoint point image must be added to the image system. In fact the potential computed by this image system gives the full analytic continuation of the potential. In this sense the image is what we will call ``reduced'', meaning the singularities occupy the smallest or a vanishingly small spatial domain. But for the spheroid, the authors of \cite{sten1995electrostatic} later realized that their expression for the image diverges when the point charge is too close to the surface \cite{lindell2001electrostatic}, and attempted to amend the image solution by subtracting an image point charge, but unlike the case of the elliptic cylinder, this proposed image system does not work - the series for the image charge density on the focal segment still diverges. The authors later studied the same problem for the \textit{dielectric} spheroid \cite{lindell2001dielectric} and used an approximate correction by extending the image from (and including) the focal segment up to the location of the point image, to imitate the line image for the dielectric sphere. 
For the non-rotational ellipsoid, the first proposed image solution consisted of a point charge plus a surface charge on an interior ellipsoid \cite{dassios2009image}. This image formulation was then presented for the particular case of the prolate spheroid in \cite{Xue2017}. But the spheroidal surface charge image cloaks a reduced image lying inside.
For the non-rotational ellipsoid with \textit{Neumann boundary condition}, the first attempt at finding an image solution is in \cite{dassios2012neumann}, where they assumed a point image plus a curved line of image charge extending along a spheroidal coordinate line, but they still needed to correct for this by adding a surface charge on an ellipsoidal surface that encloses the point and line image. This image formulation is specialized for the prolate spheroid in \cite{xue2018image}.  
A recent paper \cite{alshal2018image} suggested that the best image solution may be to use Sommerfeld images which are placed in a second copy of $\mathbb{R}^3$, attached at the surface.

There has also been a separate vein of research into the analytic continuation of solutions of more general Dirichlet boundary problems with particular attention to ellipsoids, see for example \cite{khavinson2010search}.

Here we attempt to find the reduced form of the image of a point charge near or inside a conducting spheroid. 
First in section \ref{sec 2D} we cover the 2D analogue of an elliptic cylinder, and find exact image solutions for point sources located outside and inside the ellipse. These images are used as hints for the image locations in prolate spheroids.
Section \ref{sec on axis} covers the progress made by \cite{lindell2001electrostatic} for the point charge on the rotation axis. We then devise a series expansion by geometric transformations that uncovers the true form of the image. 
In section \ref{sec off axis} we speculate about the form of the image system for a point charge located off the rotation axis, and find a simple approximate image point charge which is a generalization of the axial case, and a correction is made to a claim regarding convergence of the double series from an earlier version of the manuscript \cite{majic2021imagespheroid}.
In section \ref{sec center} we investigate the similar problem of a point charge at the center of a prolate spheroid, constructing a series solution that suggests the reduced image lies on an infinitely wide, flat sheet with a hole around the spheroid. In section \ref{sec inside} the source charge is placed anywhere inside the spheroid, and two approximate image solutions are found, depending on the location of the source. When the source is located exactly on a focal point, we find that the image is approximately represented by a multipole of order 1/2. 
%Finally, section \ref{focus} asks what it means for an image to be unique, and show a specific example where two fairly disjoint different image solutions can solve the same problem. 

%The goal here is not necessarily to find a fast and accurate approximation of a potential function, but to find its full analytic continuation by any means. With the knowledge of the location and charge distribution on the fundamental singularities, it may then be possible to construct faster and more accurate image approximations.

\section{Elliptic cylinder - 2D} \label{sec 2D}
Electrostatics problems in 2D are more suited to solving via image charges, so we study the ellipse as a precursor to the spheroid. 
%The electrostatic fields for conducting ellipses are completely solvable using image point and line sources. 
%The purpose of this section is not only to investigate image solutions for the ellipse, but these images should give us some intuitive guess for what the images may look like for the more difficult problems with spheroids later on.

Physically this problem represents a conducting elliptic cylinder in the presence of an infinite parallel line source. We will refer to the problem as if it were completely 2 dimensional, consisting of a conducting ellipse in the presence of a point charge.
 
The elliptic coordinate system is built around a focal line of length $2f$ lying on $-f\leq x\leq f$, $y=0$, where $x,y$ are Cartesian coordinates. The elliptic coordinates $\xi,\eta$ are defined as
\begin{align}
	\cosh\xi&=\frac{\sqrt{(x+f)^2+y^2}+\sqrt{(x-f)^2+y^2}}{2f}, \qquad &x&=f\cosh\xi\cos\eta, \nonumber\\
	\cos\eta&=\frac{\sqrt{(x+f)^2+y^2}-\sqrt{(x-f)^2+y^2}}{2f}, \qquad &y&=f\sinh\xi\sin\eta,
\end{align}
with $\xi\in [0,\infty)$ describing confocal ellipses around the focus and $\eta\in[0,2\pi)$ describing semi-hyperbolas. $\xi=0$ corresponds to the focal strip and $\xi$ increases with the distance from the origin $\rho=\sqrt{x^2+y^2}$. The conducting ellipse is defined by $\xi=\xi_0$.
The 2D Laplacian is separable in this coordinate system and the corresponding solutions are the
elliptic cylindrical harmonics. The interior harmonics are
\begin{align}
	\cosh(n\xi)\cos(n\eta), \qquad \sinh(n\xi)\sin(n\eta)
\end{align}
which are regular at the focus but increase exponentially as $\rho\rightarrow\infty$, so are used to expand interior potentials. And the exterior elliptic cylindrical harmonics are 
\begin{align}
	e^{-n\xi}\cos(n\eta), \qquad e^{-n\xi}\sin(n\eta)
\end{align}
which are singular on the focus but decay as $\rho\rightarrow\infty$.

\subsection{external line charge} \label{sec sten ellipse}
\begin{figure}[h]
	\includegraphics[scale=0.46]{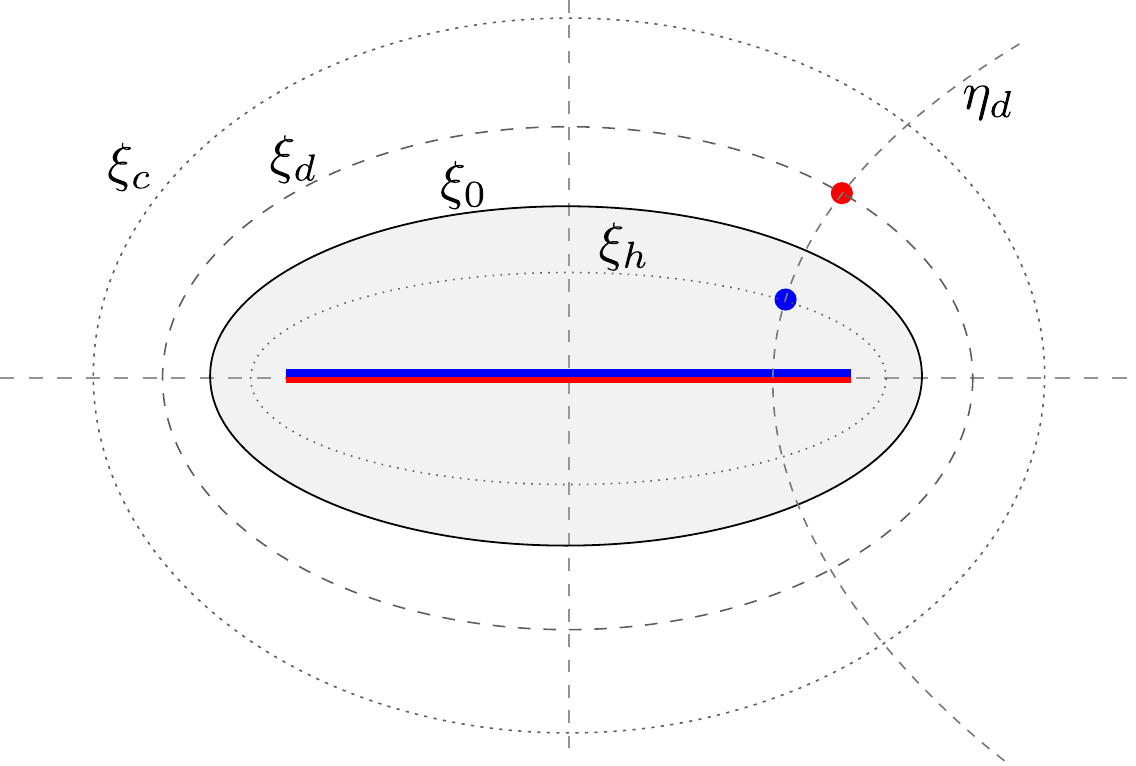}
	\caption{schematic of the image solution for a line charge near a 2d elliptic cylinder. }\label{ImageEllipticCylinder}
\end{figure}
A complete reduced image solution was derived by Sten \cite{sten1996focal} for a source outside a conducting or dielectric ellipse. This section briefly covers the derivations and describes the image for the conducting case. 

\begin{figure}
	\includegraphics[scale=.66]{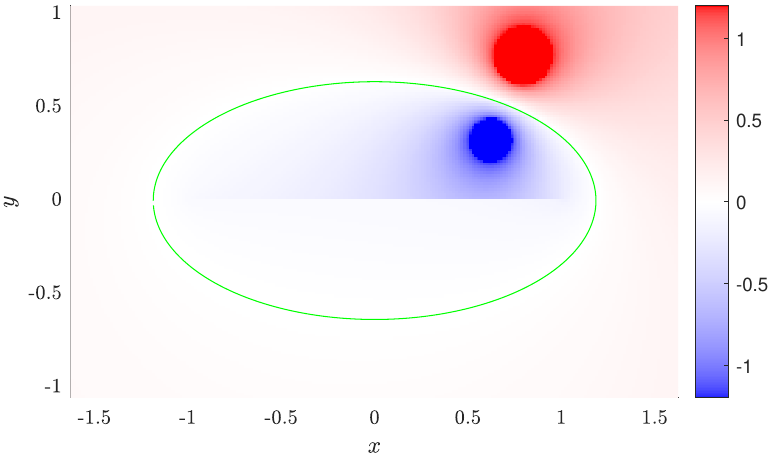}
	\caption{Analytic continuation of the potential $V_{\rm out}=V_e+V_r$ of a point source outside an ellipse (green) with $f=1$, $\xi_0=0.6$, $\xi_d=0.79$, $\eta_d=0.96$, computed via either series or the Kelvin image approach.}\label{fig Vr ellipse}
\end{figure}

The problem is to solve for the potential outside the ellipse $V$, which behaves as the point source in its vicinity, is zero on the ellipse (Dirichlet boundary condition), and decays to 0 as $\rho\rightarrow\infty$. To do this we write $V=V_e+V_r$ where $V_e$ is the known excitation - the potential of the point charge with no ellipse, and $V_r$ is the potential reflected by the ellipse which is to be solved.
The source lies at $(x,y)=(x_d,y_d)$, or $(\xi,\eta)=(\xi_d,\eta_d)$, and its potential may be expanded in elliptical harmonics for $\xi<\xi_d$ as
\begin{align}
	V_e&=\log\frac{f}{\sqrt{(x-x_d)^2+(y-y_d)^2}}\nonumber \\
	&=-\xi_0+\log 2 + \sum_{n=1}^{\infty}\frac{2}{n}e^{-n\xi_d}\big[\cosh(n\xi)\cos(n\eta_d)\cos(n\eta)+\sinh(n\xi)\sin(n\eta_d)\sin(n\eta)\big]. \label{Ve ellipse}
\end{align}
To solve for $V_r$, it is expanded as a series of exterior harmonics $e^{-n\xi}[A_n\cos(n\eta)+B_n\sin(n\eta)]$, where  $A_n$, $B_n$ are chosen to satisfy the boundary condition $V_r+V_e=0$ at $\xi=\xi_0$, giving
\begin{align}
	V_r=\xi-\xi_0+\xi_d-\log 2 - \sum_{n=1}^\infty \frac{2}{n}e^{n(\xi_0-\xi_d-\xi)}\big[\cosh(n\xi_0)\cos(n\eta_d)\cos(n\eta)+\sinh(n\xi_0)\sin(n\eta_d)\sin(n\eta)\big]. \label{Vr ellipse}
\end{align}
The terms before the sum only affect the charge of the ellipse and the zero of the potential, and have been chosen so that $V_r$ coincides with a later expression.
While the physical potential is equal to zero inside the ellipse, the series for $V_r$ converges all the way down to the focal line if the source is far enough such that $\xi_d>2\xi_0$, hence Eq. \eqref{Vr ellipse} may be identified with the potential of a charge distribution on the focal strip. The exact charge distribution may be found using the following representations of elliptical harmonics:
\begin{align}
	\xi-\log(2)=& \frac{1}{\pi}\int_{-f}^f\frac{-1}{\sqrt{f^2-x^{\prime2}}}\log\frac{f}{\sqrt{(x-x')^2+y^2}}\d x'
	\label{dist1}\\
	e^{-n\xi}\cos(n\eta)=&  \frac{n}{\pi}\int_{-f}^f\frac{T_n(x'/f)}{\sqrt{f^2-x^{\prime2}}}\log\frac{f}{\sqrt{(x-x')^2+y^2}}\d x' \qquad &n&\geq 1 \label{dist2}\\
	e^{-n\xi}\sin(n\eta)=&\frac{-1}{f\pi}\int_{-f}^f U_{n-1}(x'/f)\sqrt{f^2-x^{\prime2}}\frac{\pd}{\pd y}\log\frac{f}{\sqrt{(x-x')^2+y^2}} \d x'. \qquad &n&\geq 1 \label{dist3}
\end{align}
Eq. \eqref{dist1} is an integral over a charge distribution on the focus, concentrated towards the ends, Eq. \eqref{dist2} is a charge distribution of a Chebyshev polynomial of the first kind $T_n$, and \eqref{dist3} is a distribution of dipoles pointing in the $y$ direction weighted by a Chebyshev polynomial of the second kind $U_{n-1}$. Inserting these into Eq. \eqref{Vr ellipse} allows us to express $V_r$ as the potential of image charges and dipoles on the focal strip:
\begin{align}
	V_r=\xi_d-\xi_0 - \int_{-f}^f \bigg[ 
	q(x') \log\frac{f}{\sqrt{(x-x')^2+y^2}}+ 
	p(x')\frac{\pd}{\pd y} \log\frac{f}{\sqrt{(x-x')^2+y^2}} \bigg]\d x' \label{Vr ellipse image}	
\end{align}

where the charge distribution $q$ and dipole layer distribution $p$ are
\begin{align}
	q(x)&=\frac{f}{\pi\sqrt{f^2-x^{2}}}\left(1+2\sum_{n=0}^\infty e^{n(\xi_0-\xi_d)}\cosh(n\xi_0)\cos(n\eta_d)T_n(x/f)\right) \label{ellipse im q simple}\\
	p(x)&=-\frac{\sqrt{f^2-x^{2}}}{\pi f}\sum_{n=1}^\infty\frac{2}{n} e^{n(\xi_0-\xi_d)}\sinh(n\xi_0)\sin(n\eta_d) U_{n-1}(x/f). \label{ellipse im p simple}
\end{align}
But if the source is too close such that $\xi_d<2\xi_0$, these series diverge. To amend this, a point image source may be extracted with the correct charge and position, to make the remaining series converge all the way down to the focal strip. This image lies at  $\xi=\xi_h=2\xi_0-\xi_d$, $\eta=\eta_d$ ($x=x_h,y=y_h$) with opposite magnitude to the source, and has potential $V_h$ with the following expansion:
\begin{align}
	V_h&=\xi+\log 2-\log\frac{f}{\sqrt{(x-x_h)^2+(y-y_h)^2}} \label{Vh}\\
	&=-\sum_{n=1}^{\infty} \big[\cosh(n\xi_h)\cos(n\eta_d)\cos(n\eta)+\sinh(n\xi_h)\cos(n\eta_d)\cos(n\eta)\big]\frac{2}{n}e^{-n\xi}. \qquad \xi>\xi_h \label{Vh expansion}
\end{align}
Note that $V_h$ has zero monopole moment due to the $\xi$ term.
By adding $V_h$ in the form \eqref{Vh} to $V_r$ and subtracting the form \eqref{Vh expansion}, we get a series for $V_r$ that converges down to the focal strip:

\begin{align}
	V_r(\xi_d>2\xi_0)= ~&\xi_d-\xi_0-\log\frac{f}{\sqrt{(x-x_h)^2+(y-y_h)^2}}\nonumber\\
 &-\sum_{n=1}^{\infty}\frac{2}{n}\bigg(\big[e^{n(\xi_0-\xi_d)}\cosh(n\xi_0)-\cosh(n\xi_h)\big]\cos(n\eta_d)\cos(n\eta) \nonumber\\
	 &\hspace{1cm}+ \big[e^{n(\xi_0-\xi_d)}\sinh(n\xi_0)-\sinh(n\xi_h)\big]\sin(n\eta_d) \sin(n\eta)\bigg)e^{-n\xi}
\end{align}
which corresponds to the image system if $\xi_d\leq2\xi_0$:
\begin{align}
	q_h(x,y)&=-\delta(x-x_h)\delta(y-y_h) \label{ellipse im h}\\
	q(x)&=
	\frac{2f}{\pi\sqrt{f^2-x^{2}}}\sum_{n=1}^\infty \big[e^{n(\xi_0-\xi_d)}\cosh(n\xi_0)-\cosh(n\xi_h)\big]\cos(n\eta_d)T_n(x/f) \label{ellipse im q}\\
	p(x)&=-\frac{\sqrt{f^2-x^{2}}}{\pi f}\sum_{n=1}^\infty\frac{1}{n} \big[e^{n(\xi_0-\xi_d)}\sinh(n\xi_0)-\sinh(n\xi_h)\big]\sin(n\eta_d) U_{n-1}(x/f) \label{ellipse im p}
\end{align}
These images are depicted schematically in Figure \ref{ImageEllipticCylinder}, and the potential is plotted in Figure \ref{fig Vr ellipse}. The potential of these images is the analytic continuation of $V_r$ to all space. 
%The dipole layer causes the sharp discontinuity across the focal segment.

In two-dimensions analytic continuation of solutions to Dirichlet problems is much more tractable for many geometries, using a wide range of conformal mappings, or the Schwarz function \cite{bell2006classical}. In the next section we look at an elegant approach using reflection \cite{alshal2021image} which provides a closed form expression for the image.

\subsection{Alternate method} \label{sec alshal ellipse}
The same problem of a point source outside a conducting ellipse is solved in Ref. \cite{alshal2021image} via the Kelvin and Sommerfeld image methods, and these methods give simple expressions for the analytic continuation of the potential. The two methods essentially obtain the same expression, so we just reproduce the Kelvin method here.

The approach uses a function $G$, similar to Green's function of an isolated point source, but where $\mathbb{R}^2$ is replaced by a double surface attached at the focal line \cite{curtright2018conducting}:
\begin{align}
	G(\xi,\eta,\xi_d,\eta_d)=-\frac{1}{2}|\xi-\xi_d| - \frac{1}{2}\log\big(1-e^{-2|\xi-\xi_d|}+2e^{-|\xi-\xi_d|}\cos(\eta-\eta_d)\big). \label{G}
\end{align}
$G$ may also be viewed as the potential of a point charge plus a charge distribution on the focal strip.
To construct a solution $V$ to Laplace's equation that is zero on the boundary of the ellipse $\xi_0$, one simply takes the difference of two offset copies of $G$ as
\begin{align}
	V=G(\xi,\eta,\xi_d,\eta_d)-G(\xi,\eta,\xi_h,\eta_d). \label{V ellipse Kelvin}
\end{align}
where again $\xi_h=2\xi_0-\xi_d$ is the $\xi$ parameter of the image point source (note $\xi_h<0$ is allowed and corresponds to the point source lying in the second copy of $\mathbb{R}^2$).
Note that the first term in \eqref{V ellipse Kelvin} is not just the source and the second term is not the complete image, because both contribute towards image charges on the focal segment. To be explicit, the reflected potential is
$V_r=V-V_e$, which gives the closed form of the series solution \eqref{Vr ellipse}.

\begin{figure}
	\includegraphics[scale=.625]{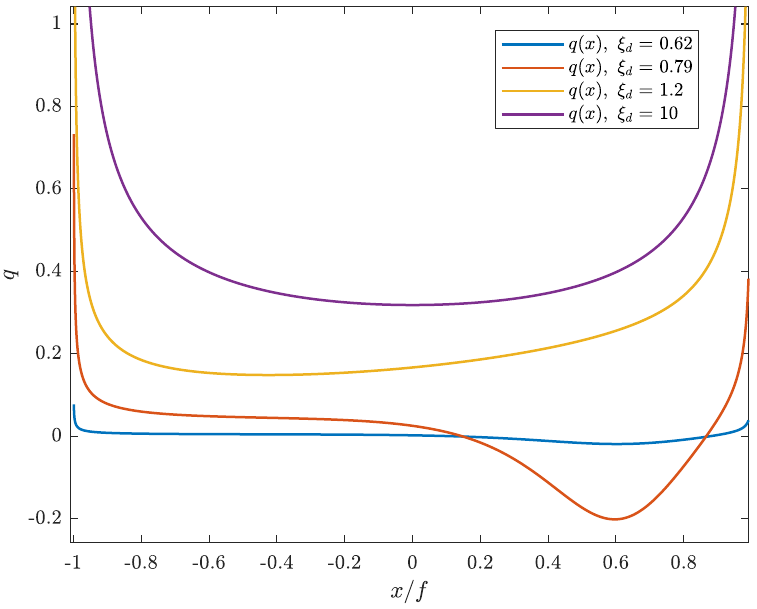}
	\includegraphics[scale=.625]{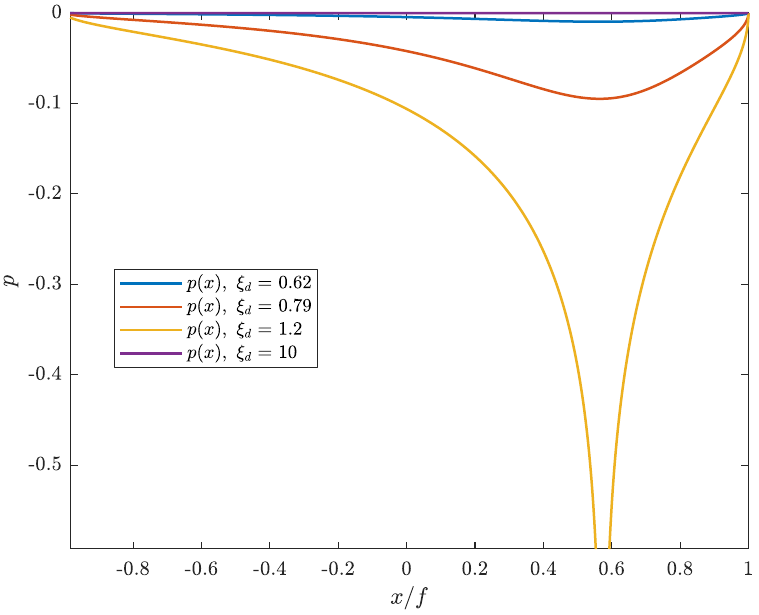}
	\caption{Charge density $q$ (left) and dipole density $p$ (right) for a 2D point charge outside an ellipse with $\xi_0=0.6$, $\eta_d=.96$ (same as in Fig. \ref{fig Vr ellipse}), and a range of $\xi_d$ including $\xi_d=2\xi_0$.  }\label{fig ellipse q(x)}
\end{figure}

A closed form of the image charge and dipole distributions can be found by evaluating the electric field at the focal strip. 
The image charge distribution causes a discontinuity in the electric field crossing the focal strip:
\begin{align}
	\left.\frac{\pd V}{\pd y}\right|_{y\rightarrow0^+}-\left.\frac{\pd V}{\pd y}\right|_{y\rightarrow0^-}=2\pi q(x) 
\end{align}
And the image dipole layer causes a discontinuity in the potential across the focal strip:
\begin{align}
	V(x,y\rightarrow0^+)-V(x,y\rightarrow 0^-)=2\pi p(x) \label{dipole layer}
\end{align}
Using these with the potential Eq. \eqref{V ellipse Kelvin} gives alternative but equivalent expressions to those in Eqs. (\ref{ellipse im p simple},\ref{ellipse im q simple}) if $\xi_d\geq2\xi_0$, or (\ref{ellipse im p},\ref{ellipse im q}) if $\xi_d<2\xi_0$:
\begin{align}
q(x)=&\frac{f}{4\pi\sqrt{f^2-x^2}}	
	\bigg[\frac{\sinh\xi_d}{\cosh\xi_d-\cos(\eta_d+\text{acos}(x/f))}
          -\frac{\sinh\xi_h}{\cosh\xi_h-\cos(\eta_d+\text{acos}(x/f))}\nonumber\\
	     &\hspace{2cm}+\frac{\sinh\xi_d}{\cosh\xi_d-\sin(\eta_d+\text{asin}(x/f))}
          -\frac{\sinh\xi_h}{\cosh\xi_h-\sin(\eta_d+\text{asin}(x/f))}\bigg] \\
 p(x)=&\frac{1}{4\pi}	
 \bigg[\log\frac{\cosh\xi_d +\sin(\eta_d-\text{asin}(x/f))}{\cosh\xi_d -\sin(\eta_d+\text{asin}(x/f))} 
       -\log\frac{\cosh\xi_h +\sin(\eta_d-\text{asin}(x/f))}{\cosh\xi_h -\sin(\eta_d+\text{asin}(x/f))}\bigg]  .    
\end{align}
These expressions do not depend on whether the image point source is present (i.e. if $\xi_d<2\xi_0$), despite the absolute value signs $|\xi-\xi_h|$ appearing in \eqref{V ellipse Kelvin}. Figure \ref{fig ellipse q(x)} shows that $q(x)$ varies smoothly as $\xi_d$ crosses $2\xi_0$, while $p(x)$ has an unbounded peak for $\xi_d=2\xi_0$.\\

Ref. \cite{alshal2021image} suggests that this type of image solution could generalize to spheroids, although surely the generalization would be much more complicated, and we can't reproduce the result.

\subsection{Source at center} \label{sec ellipse center}
\begin{figure}
	\includegraphics[scale=.7]{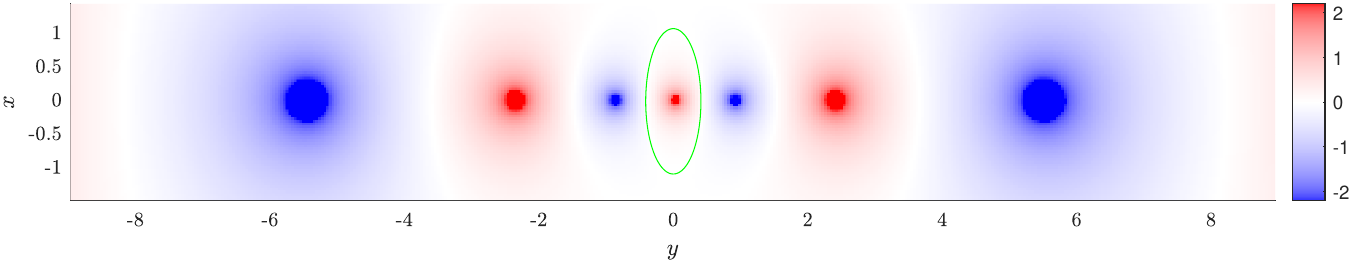}
	\caption{Analytic continuation of the potential of a point source at the center of an ellipse (green) with $f=1$, $\xi_0=0.4$.}\label{fig ellipseimagecenter}
\end{figure}

The  problem of a point source inside a conducting ellipse appears to have attracted no attention. 
So here we solve the problem and determine the image outside the ellipse, starting with the simple case of a point source placed at the origin. The potential of the line charge should be expanded as
\begin{align}
	V_e&=\log\frac{f}{\rho} \nonumber\\
	&= -\xi + \log 2 + \sum_{n=1}^\infty \frac{2}{n} e^{-n\xi}\cos\frac{n\pi}{2}\cos n\eta \label{Ve ellipse center}
\end{align}
and the reflected potential is expanded on a series of $\cosh n\xi\cos(n\eta)$ which are smooth inside the ellipse:
\begin{align}
	V_r=\xi_0-\log 2 -\sum_{n=1}^\infty  \frac{2}{n}\frac{e^{-n\xi_0}}{\cosh(n\xi_0)}\cosh(n\xi) \cos\frac{n\pi}{2}\cos n\eta.  \label{Vr ellipse center}
\end{align}
 Just like for the external point source, we want to extract the most significant contributions (leading order with respect to $n$) from the series. In fact the following series expansion gives all orders in $n$:
\begin{align}
	\frac{e^{-n\xi_0}}{\cosh(n\xi_0)} =& -2e^{-2n\xi_0}+2e^{-4n\xi_0}-2e^{-6n\xi_0}+... \nonumber\\
	=&2\sum_{k=1}^\infty (-)^k e^{-2kn\xi_0} \qquad \xi_0>0. \label{expn k}
\end{align}
The sum over $n$ of the $k^{th}$ term gives two identical point sources at $x=0,~y_k=\pm \sinh(2k\xi_0)$ (plus a constant term), so $V_r$ can be represented as a series of point charges of alternating sign whose spacing increases with distance from the ellipse:
\begin{align}
	V_r=\xi_0-\log 2 + \sum_{k=1}^\infty (-)^k \bigg( 4k\xi_0 - 2\log 2 + \log\frac{f}{\sqrt{x^2+(y-y_k)^2}} + \log\frac{f}{\sqrt{x^2+(y+y_k)^2}}\bigg). \label{Vr ellipseimagecenter}
\end{align}
The analytic continuation of the potential $V_e+V_r$ is plotted in Fig. \ref{fig ellipseimagecenter} using Eq. \eqref{Vr ellipseimagecenter}, revealing the image charges. 
The image problem of a point charge at the center of a spheroid is much more complex as seen in section \ref{sec center}.

\subsection{Internal source off center}\label{sec ellipse off center}
This problem is also solvable with a relatively simple image system. The excitation is
\begin{align}
	V_e&=\log\frac{f}{\sqrt{(x-x_d)^2+(y-y_d)^2}}\nonumber \\
	&=-\xi+\log 2 + \sum_{n=1}^{\infty}\frac{2}{n}e^{-n\xi}\big[\cosh(n\xi_d)\cos(n\eta_d)\cos(n\eta)+\sinh(n\xi_d)\sin(n\eta_d)\sin(n\eta)\big] \label{Ve ellipse inside}
\end{align}
and $V_r$ is found to be
\begin{align}
	V_r=\xi_0-\log 2 - \sum_{n=1}^\infty \frac{2}{n}e^{-n\xi_0}\bigg[\frac{\cosh(n\xi_d)}{\cosh(n\xi_0)}\cosh(n\xi)\cos(n\eta_d)\cos(n\eta)
	+\frac{\sinh(n\xi_d)}{\sinh(n\xi_0)}\sinh(n\xi)\sin(n\eta_d)\sin(n\eta)\bigg]. \label{Vr ellipse inside}
\end{align}
Here we can expand the series coefficients as
\begin{align}
	e^{-n\xi_0}\frac{\cosh(n\xi_d)}{\cosh(n\xi_0)} &= -(e^{n\xi_d}+e^{-n\xi_d})\sum_{k=1}^{\infty}(-)^k e^{-2kn\xi_0} \nonumber\\
	e^{-n\xi_0}\frac{\sinh(n\xi_d)}{\sinh(n\xi_0)} &= (e^{n\xi_d}-e^{-n\xi_d})\sum_{k=1}^{\infty} e^{-2kn\xi_0} 
\end{align}
A reasonable guess for the image system would be a series of point charges similar to those found in the previous section. Towards associating the series with expansions of point sources we organize the series for $V_r$ in to two parts, with $e^{n\xi_0}$ or $e^{-n\xi_0}$:
\begin{align}
V_r=\xi_0-\log 2 + \sum_{k=1}^\infty \bigg\{
	 &\sum_{n=1}^\infty \frac{2}{n}e^{-n(2k\xi_0+\xi_d)}\big[(-)^k\cosh(n\xi)\cos(n\eta_d)\cos(n\eta)
	-\sinh(n\xi)\sin(n\eta_d)\sin(n\eta)\big] \nonumber\\
	+&\sum_{n=1}^\infty \frac{2}{n}e^{-n(2k\xi_0-\xi_d)}\big[(-)^k\cosh(n\xi)\cos(n\eta_d)\cos(n\eta)
	+\sinh(n\xi)\sin(n\eta_d)\sin(n\eta)\big]\bigg\}. \label{Vr ellipse inside nk}
\end{align}
By adding constants, these can be identified with expansions of point sources at $(\xi,\eta)=(2k\xi_0\pm\xi_d,\pm\eta_d)$, leading to the image formulation
\begin{align}
	V_r=\sum_{k=1}^\infty (-)^k \bigg[4k\xi_0 -2\log2 + \log\frac{f}{\sqrt{(x-x_{k+})^2+(y-y_{k+})^2}}
	+ \log\frac{f}{\sqrt{(x-x_{k-})^2+(y-y_{k-})^2}} \bigg] \label{Vr ellipse inside image}
\end{align}
where the coordinates of the image point charges are
\begin{align}
	x_{k\pm}=f\cosh(2k\xi_0\pm\xi_d)\cos\eta_d , \qquad y_{k+\pm}=f\pm(-)^k\sinh(2k\xi_0\pm\xi_d)\sin\eta_d. \label{xkyk}
\end{align}
\begin{figure}
	\includegraphics[scale=.7]{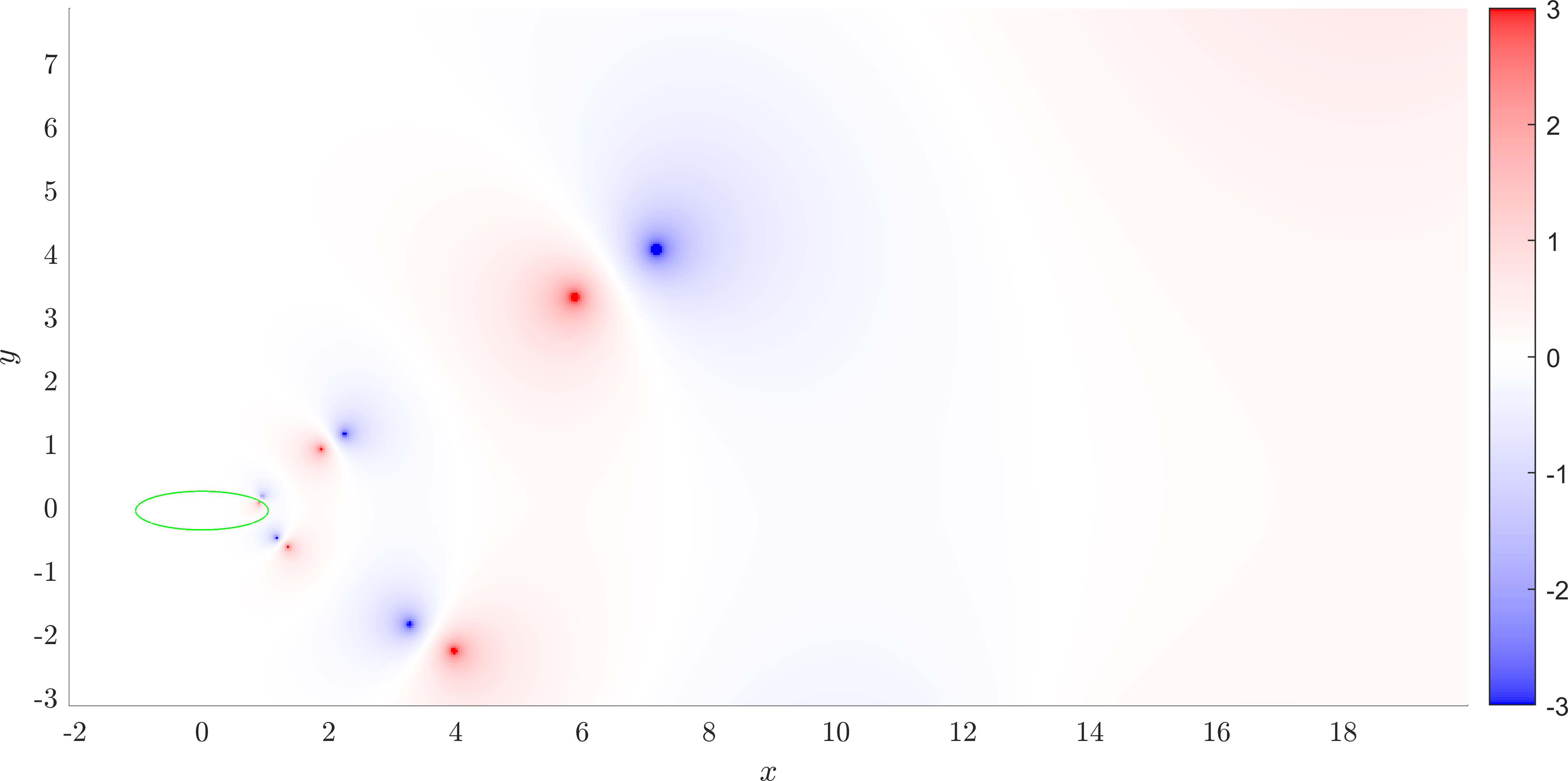}
	\caption{Analytic continuation of the potential inside an ellipse with $f=1$, $\xi_0=0.4$, and point source at $\xi_d=0.3,~\eta_d=\pi/6$.} \label{fig ellipseimageinside}
\end{figure}

\begin{figure}
	\includegraphics[scale=.7]{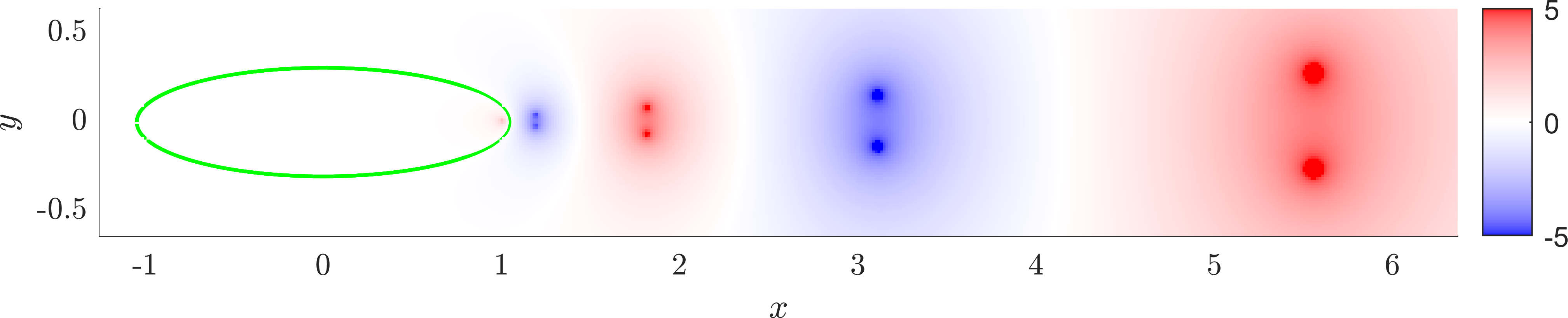}
	\caption{Analytic continuation of the potential inside an ellipse with $f=1$, $\xi_0=0.3$, and point source at $\xi_d=0,~\eta_d=0.05$.} \label{fig ellipseimageinsidetip}
\end{figure}

\begin{figure}
	\includegraphics[scale=.7]{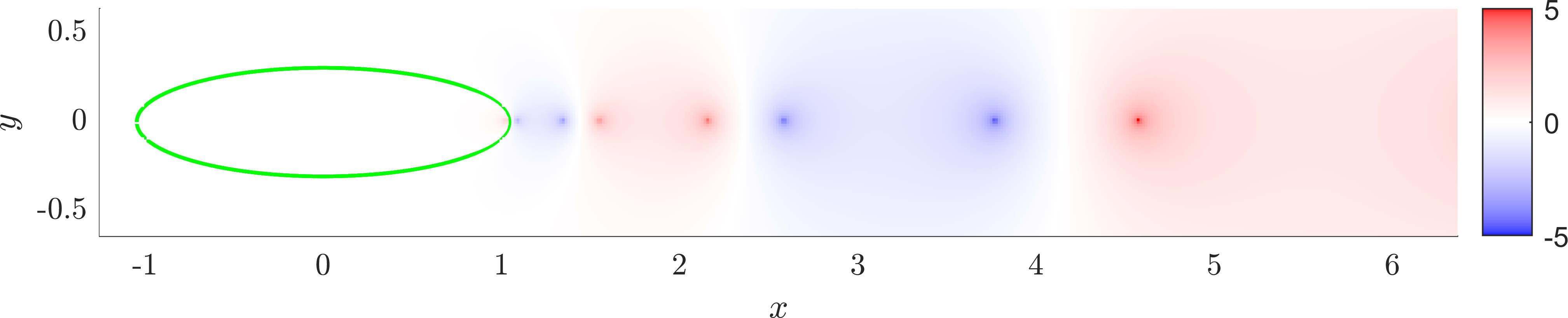}
	\caption{Analytic continuation of the potential inside an ellipse with $f=1$, $\xi_0=0.3$, and point source at $\xi_d=0.2,~\eta_d=0$.} \label{fig ellipseimageinsidetiptip}
\end{figure}

The analytic continuation of the total potential is plotted in Figure \ref{fig ellipseimageinside} using Eq. \eqref{Vr ellipse inside image}. The point charges lie on the hyperbola $\cos\eta=\cos\eta_d$ with alternating sign and alternating spacing that increases exponentially going further out. Numerically the total potential is also zero on infinitely many larger confocal ellipses.
As $\xi_d\rightarrow0, \eta_d\rightarrow\pi/2$ the image reduces to that for the source at the center, studied in the previous section. As the ellipse becomes more circular, all except one image charge move out towards infinity, recovering the image solution for the circle. 
For a charge off center but on the focal strip,  the images group in twos either side of the $x$-axis as shown in figure \ref{fig ellipseimageinsidetip}. 
For a charge off the focal strip, on the $x$-axis near the tip, the images combine and line up on the $x$-axis as shown in figure \ref{fig ellipseimageinsidetiptip}.

This concludes the 2D section of the paper, for which we could find the full analytic continuation in every case. 
These image systems should provide hints for the image locations in the 3D problem in the next sections.

\section{Point charge on axis of conducting prolate spheroid} \label{sec on axis}
Consider a conducting prolate spheroid of half-focal length $f$, half-height $c$ and half-width $a=\sqrt{c^2-f^2}$, where in Cartesian coordinates $x,y,z$ the rotation axis is aligned with $z$. The spheroid is excited by a point charge located a distance $d$ up the $z$-axis, as depicted in Figure \ref{SpheroidLineImage}. 
We will use spheroidal coordinates $\xi,\eta$, with $\rho=\sqrt{x^2+y^2}$, defined as: \cite{NIST:DLMF}

Spheroidal coordinates $\xi,\eta$ are analogous to elliptic coordinates, but here we do not use the trigonometric representations:
\begin{align*}
\xi&=\frac{\sqrt{\rho^2+(z+f)^2}+\sqrt{\rho^2+(z-f)^2}}{2f}, & z&=f\xi\eta,\\
\eta&=\frac{\sqrt{\rho^2+(z+f)^2}-\sqrt{\rho^2+(z-f)^2}}{2f},& \rho&=f\sqrt{\xi^2-1}\sqrt{1-\eta^2} 	.
\end{align*}
Surfaces of constant $\xi$ are concentric prolate spheroids, with $\xi=1$ being the focal line, and surfaces of constant $\eta$ are hyperboloids which describe an analogue of the latitudinal angle on a given spheroid. 
The boundary of the spheroid is $\xi=\xi_0=c/f$. 
The exciting potential $V_{e}$ can be expressed as a series of spheroidal harmonics:
\begin{align}
V_{e}=\frac{f}{\sqrt{\rho^2+(z-d)^2}} =\sum_{n=0}^\infty (2n+1) Q_n(\xi_d) P_n(\xi)P_n(\eta).  \qquad \xi<\xi_d\label{Ve axis}
\end{align}
Where $P_n$ and $Q_n$ are the Legendre functions of the first and second kinds, and $\xi_d=d/f$ is the $\xi$ coordinate of the source charge.
We split the total potential outside the spheroid as  $V=V_{e}+V_{r}$ where $V_{r}$ is the reflected potential\footnote{the terminology ``reflected potential" is not intended to suggest the existence of a reflection law as for example with Kelvin's method for the sphere.} by the presence of the spheroid, and the boundary condition is then $V=0$ at $\xi=\xi_0$.
Using the standard separation of variables approach, $V_{r}$ is found to be
\begin{align}
V_{r}=-\sum_{n=0}^\infty (2n+1) \frac{P_n(\xi_0)}{Q_n(\xi_0)}Q_n(\xi_d) Q_n(\xi)P_n(\eta). \label{Vr}
\end{align}
We will now investigate the analytic continuation of the series \eqref{Vr} inside the spheroid.

\subsection{Summary of progress in Ref. \cite{lindell2001electrostatic} - leading order approximation and series acceleration} \label{sec lindell}
\begin{figure}
	\centering{\includegraphics[scale=0.4]{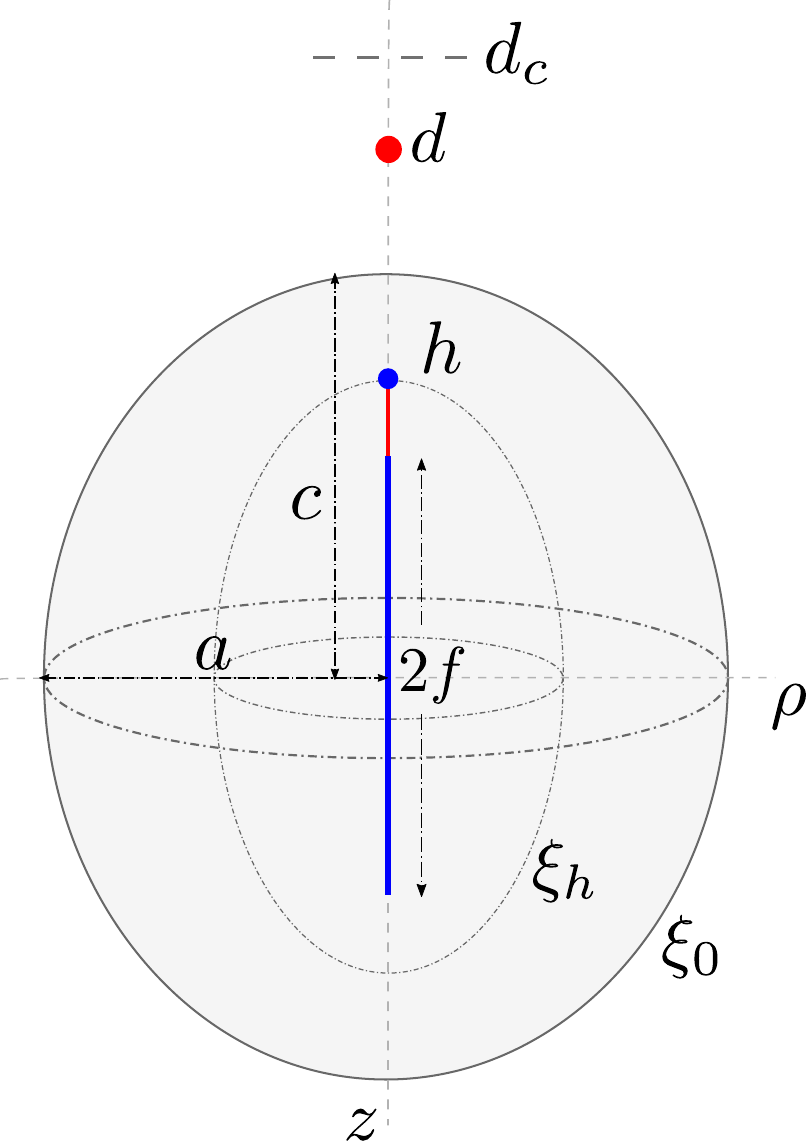}}
	\caption{Schematic of a point charge near a prolate spheroid, showing the extended image singularity in the case of a close source charge, and the boundary of the domain of convergence of the series \eqref{Vr}, $\xi=\xi_h$. Red represents positive charge and blue negative.}\label{SpheroidLineImage}
\end{figure}
%We will first briefly cover the results of \cite{lindell2001electrostatic}. 
%, where they discovered that the image singularity of a point charge on the axis of a prolate spheroid changes form when the point charge is moved past a critical distance from the surface. 
The Legendre functions behave for large $n$ with $m$ fixed, $x>1$ as
\begin{align}
P_n^m(x)\rightarrow&~\sqrt{\frac{1}{2\pi}} \frac{(x+\sqrt{x^2-1})^{n+1/2}}{(x^2-1)^{1/4}}n^{m-1/2}  \left[1-\frac{2m-1}{8n}\left( \frac{(2m+1)x}{\sqrt{x^2-1}}-2\right)+\cO(n^{-2})\right] \qquad~~ |x|>1\\
Q_n^m(x)\rightarrow&~\sqrt{\frac{\pi}{2}}\frac{(x+\sqrt{x^2-1})^{-n-1/2}}{(x^2-1)^{1/4}}n^{m-1/2} \left[1+\frac{2m-1}{8n}\left( \frac{(2m+1)x}{\sqrt{x^2-1}}+2 \right)+\cO(n^{-2})\right]\qquad~~ |x|>1 \label{lim legendre}
\end{align}
which are modified expressions from \cite{nemes2020large}. For now we are just interested in the leading order and $m=0$, where the terms in the series \eqref{Vr} behave for large $n$ as
\begin{align}
\frac{P_n(\xi_0)}{Q_n(\xi_0)}Q_n(\xi_d)=\frac{1}{\sqrt{2\pi n}(\xi_d^2-1)^{1/4}}\left(	\frac{\Big(\xi_0+\sqrt{\xi_0^2-1}\Big)^2}{\xi_d+\sqrt{\xi_d^2-1}}\right)^n\big[1+\cO(n^{-1}) \big]. \label{lim lindell}
\end{align} 
If the point source is further than a critical distance, $d>d_c$ where
\begin{align}
d_c = 2 \frac{c^2}{f}-f=f(2\xi_0^2-1), \label{dc}
\end{align}

then the series \eqref{Vr} converges in all space except the focal segment.
%, and the corresponding image may be expressed as a series using the Havelock formula for the prolate spheroidal harmonics.
The corresponding image line charge distribution $\lambda$ can be found by applying the Havelock formula \cite{havelock1952moment}(p. 130) which expresses each spheroidal harmonic in terms of a charge distribution of  a Legendre polynomial on the focal segment:
\begin{align}
	Q_n(\xi)P_n(\eta)=\frac{1}{2}\int_{-f}^f\frac{P_n(z')}{\sqrt{\rho^2+(z-z')^2}} \d z'.
\end{align}
Combining this with Eq. \eqref{Vr} reveals the line charge density of the image:
\begin{align}
\lambda(z)= -\frac{1}{2}\sum_{n=0}^\infty (2n+1) \frac{P_n(\xi_0)}{Q_n(\xi_0)}Q_n(\xi_d) P_n(z/f), \label{varrho} \qquad |z|<f
\end{align}
so that the reflected potential may be expressed in its image representation as
\begin{align}
	V_r=\int_{-f}^{f} \frac{\lambda(z')}{\sqrt{\rho^2+(z-z')^2}}\d z'
\end{align}
which converges if $d > d_c$. But for $d<d_c$, the series \eqref{Vr} diverges inside the spheroid $\xi<\xi_h$ where
\begin{align}
\xi_h=(2\xi_0^2-1)\xi_d - 2\xi_0\sqrt{\xi_0^2-1}\sqrt{\xi_d^2-1}. \label{xih}
\end{align} 
%This is shown schematically in Figure \ref{SpheroidLineImage}.
%, where the spheroidal series diverges inside the inner spheroid. 

To work around this, the authors noted that the leading order of the series terms in \eqref{lim lindell} is similar to that for the spheroidal harmonic expansion of a point charge on the z-axis at $z=h=f\xi_h$: 
\begin{align}
V^{(0)}=&\frac{q_h f}{\sqrt{\rho^2+(z-h)^2}}  \label{V^(0)}\\
=&q_h\sum_{n=0}^\infty (2n+1) P_n(\xi_h) Q_n(\xi)P_n(\eta), \label{V^(0)sum}\\
\text{where }~~ q_h=&-\sqrt[4]{\frac{h^2-f^2}{d^2-f^2}}. \nonumber
\end{align}  
This is the blue dot in Figure \ref{SpheroidLineImage}. Then \eqref{V^(0)sum} was extracted from the series \eqref{Vr} so that the remaining series converged faster:
\begin{align}
V\equiv V'=V_{e}+V^{(0)}-\sum_{n=0}^\infty (2n+1) \bigg(\frac{P_n(\xi_0)}{Q_n(\xi_0)}Q_n(\xi_d) + q_h P_n(\xi_h)\bigg) Q_n(\xi)P_n(\eta). \label{V'}
\end{align} 
However, the authors incorrectly assumed that the remaining image lies on the focal segment, and attempted to write down the image charge density on this strip for $d<d_c$ as:
\begin{align}
	\lambda(z)\neq - \frac{1}{2}\sum_{n=0}^\infty (2n+1)  \bigg(\frac{P_n(\xi_0)}{Q_n(\xi_0)}Q_n(\xi_d) + q_h P_n(\xi_h)\bigg) P_n(z/f), \label{varrho neq} \qquad |z|<f
\end{align}
but the terms increase exponentially as $n\rightarrow\infty$ and the series diverges. Ref. \cite{lindell2001electrostatic} states that this series is asymptotic, initially converging then diverging, but no charge distribution confined to this segment could solve the boundary problem. 
%The analytic continuation of the potential is actually singular in the region off this focus too on the segment $f<z<h$, (see Section \ref{sec on axis}), and no confined charge distribution can produce a potential with a singularity outside the location of the charges. 
In order for the potential created by this charge distribution on the focal segment to equal $V_r$ in any open region, it must be equal to $V_r$ $everywhere$ due to the identity theorem \cite{lebl2019tasty}. This is not possible because $V_r$ also tends to infinity near $\rho=0, ~f<z<h$ (see Section \ref{analytic}) and this behavior cannot be reproduced by any charge distribution confined to $-f\leq z\leq f$.\\

%Perhaps one  could argue that the image is still valid in some sense because it can be represented by a series of spheroidal harmonics which converges for $\xi>\xi_h$, but then nothing is gained from this ``image" representation since it is only useful to reproduce the series it was obtained from in the first place.
If $d>d_c$ then the image point charge lies off the focal segment, and is in a sense absorbed into the image line charge distribution $\lambda(z)$. As noted in Ref. \cite{lindell2001electrostatic}, extracting the image charge $V^{(0)}$ only improves the rate of convergence if $d>d_c$.  
%\footnote{The authors actually state that the image approximation requires $h\geq f$ (that the image charge lies off the focal segment) which in turn is equivalent to $d<d_c$. But in fact $h\geq f$ regardless of whether $d<d_c$, as shown in Figure \ref{CompareImageLocations}). 
%%	Although it can happen that $h>c$ for some much larger $d$. 
%%Nevertheless, there is no point in extracting an image point charge at $z=h>f$ when the image lies entirely on the focal segment anyway, because the extraction actually makes the series \eqref{V'} converge slower. 
%To convert to the notation of \cite{lindell2001electrostatic}, set $a\rightarrow b$ $c\rightarrow a$, $f\rightarrow c$, $d_c\rightarrow d_o$.}

\subsection{Comparison to approach in Ref. \cite{Xue2017}}

\begin{figure}
	\centering{\includegraphics[scale=0.7]{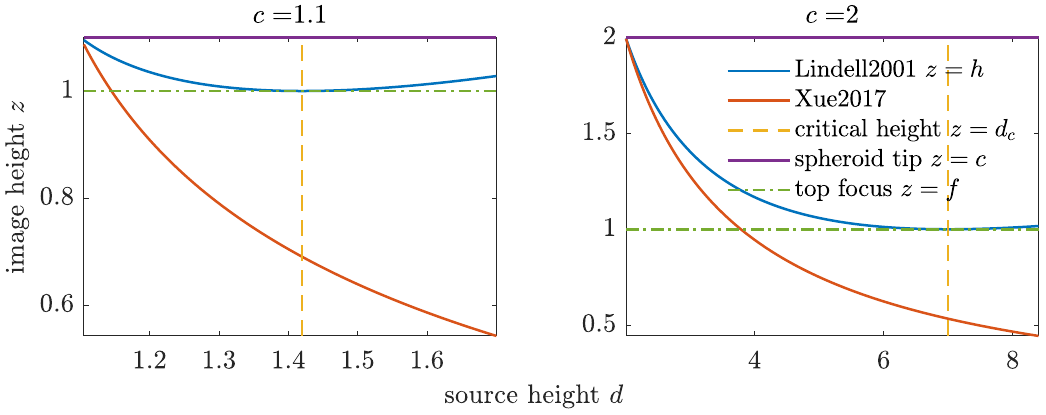}\\
		\includegraphics[scale=0.7]{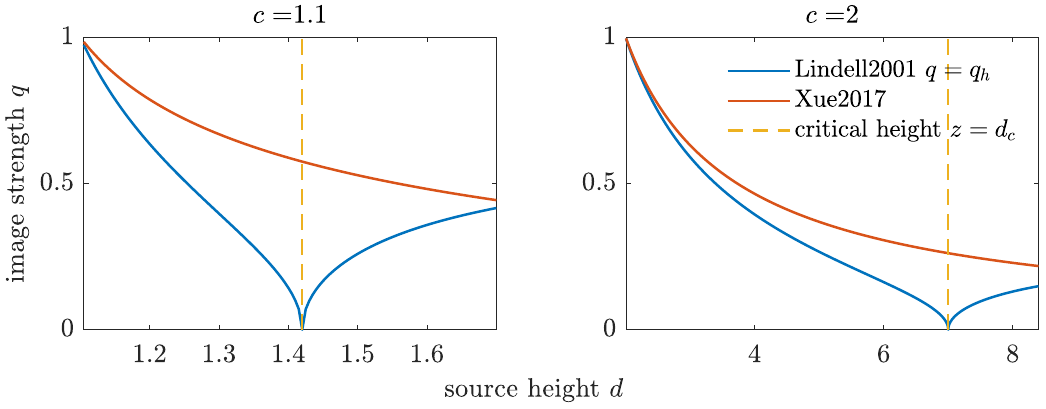}}
	\caption{Comparison of the height $h$ of the image point charge in \cite{lindell2001electrostatic} to that from \cite{Xue2017}. The height is plotted as a function of the source charge height $d$, for two spheroids of very different aspect ratio. The top of the focal segment lies at $f=1$, and $d_c$ is the critical height for the source charge - the blue line is only used for $d<d_c$.} \label{CompareImageLocations}
\end{figure}
In Ref. \cite{Xue2017} the authors separated out a different image point charge, where its location and strength were chosen so as to cancel out just the $n=0$ and $n=1$ terms of the series \eqref{Vr} exactly. 
They used for the height and charge:
\begin{align}
	h_x=c\frac{Q_0(\xi_0)Q_1(\xi_d)}{Q_1(\xi_0)Q_0(\xi_d)},\qquad
	q_x= c \frac{Q_0(\xi_d)}{Q_0(\xi_0)}. \label{hxqx}
\end{align}
As shown in Fig. \ref{CompareImageLocations}, the height of the image $h_x$ in Ref. \cite{Xue2017} decreases all the way to the center of the spheroid as the source moves away from the surface (as $d$ increases), unlike $h$ from Ref. \cite{lindell2001electrostatic} which avoids the focal segment. And the charge $q_x$ decreases smoothly as $d$ increases, and does not recognize the critical distance $d=d_c$ as does $q_h$. The image approximation \eqref{hxqx} applies well in cases where the spheroidal series converges very quickly so that the first two terms are dominant, which tends to be the case for more distant sources and rounder spheroids, but from the point of view of analytical continuation, this point charge does not match the singularity of the potential.

%Now we shall look at one method of analytically continuing $V'$ to look at the true nature of the image.

\subsection{Analytic continuation of the potential for close source charges }\label{analytic}

For point sources within the critical distance $d<d_c$, the potential diverges within the spheroid $\xi<\xi_h$, which extends along the $z$-axis to $z=\pm h$. We will make the intuitive guess that $V$ is singular on the $z$-axis from $z=-f$ to $h$, shown in Figure \ref{SpheroidLineImage}; there seems no reason why the image would have to extend to $z<-f$, when the image should concentrate towards the top surface as the source comes very close, mimicking the image solution for the plane. And it is a fair assumption that the image lies on the axis, as is the case for $d>d_c$.

So we'll define a `stretched' prolate coordinate system whose focal segment lies on the $z$-axis from $z=-f$ to $h$, exactly on the proposed singularity:
\begin{align}
\bar\xi&=\frac{\sqrt{\rho^2+(z+f)^2}+\sqrt{\rho^2+(z-h)^2}}{f+h} \\
\bar\eta&=\frac{\sqrt{\rho^2+(z+f)^2}-\sqrt{\rho^2+(z-h)^2}}{f+h},
\end{align}
with the goal of expressing $V_{r}$ as a series of the corresponding spheroidal harmonics $Q_n(\bar\xi)P_n(\bar\eta)$. If $V_r$ is only singular on $\rho=0,~-f\leq z\leq h$, such a series should converge everywhere except on this singularity.
\begin{figure}
		\includegraphics[scale=0.62,trim={0 5mm 0 4mm}, clip]{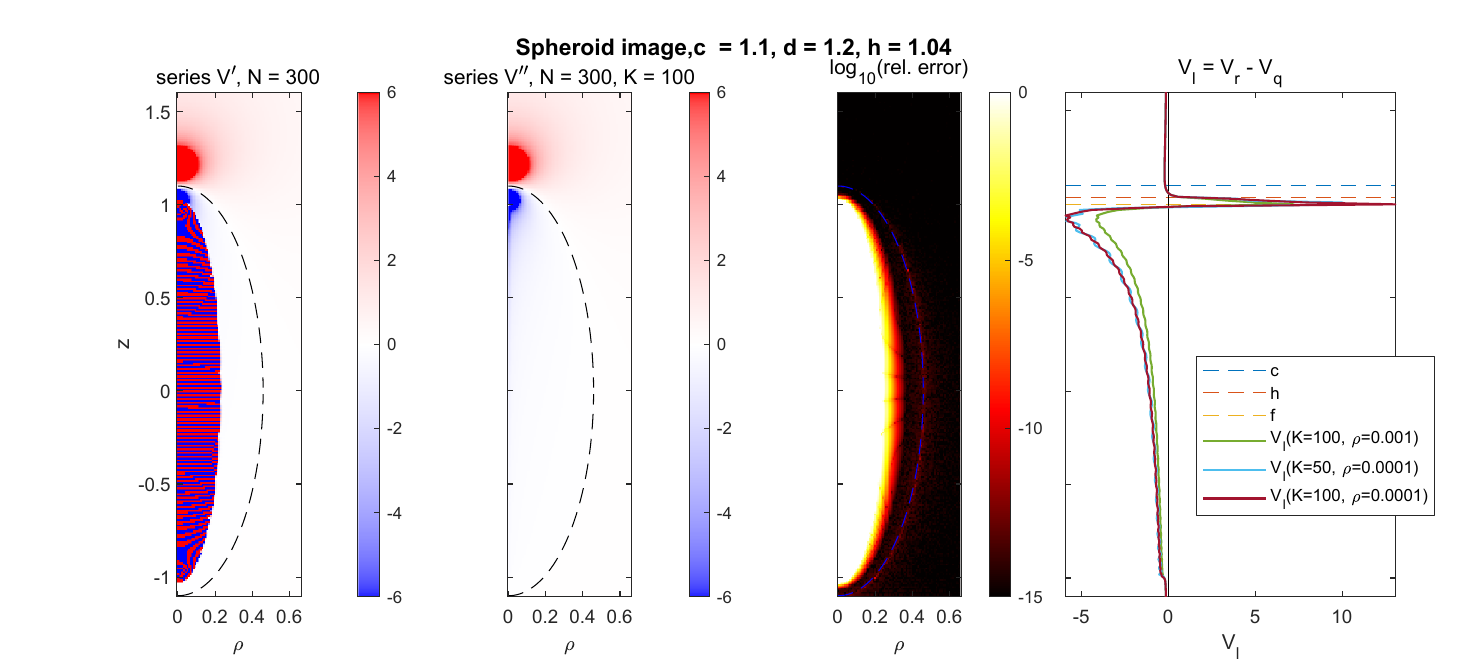}
		\includegraphics[scale=0.62,trim={0 5mm 0 4mm}, clip]{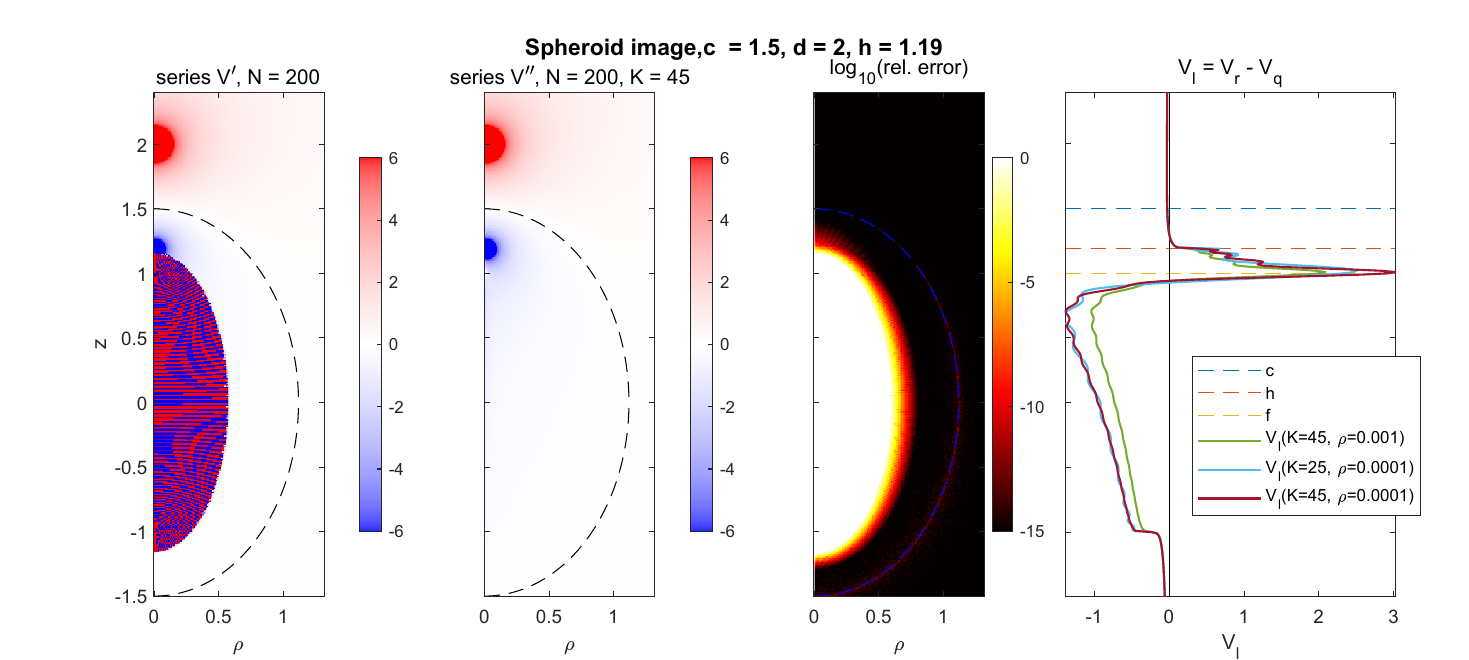}
		\includegraphics[scale=0.62,trim={0 0mm 0 4mm}, clip]{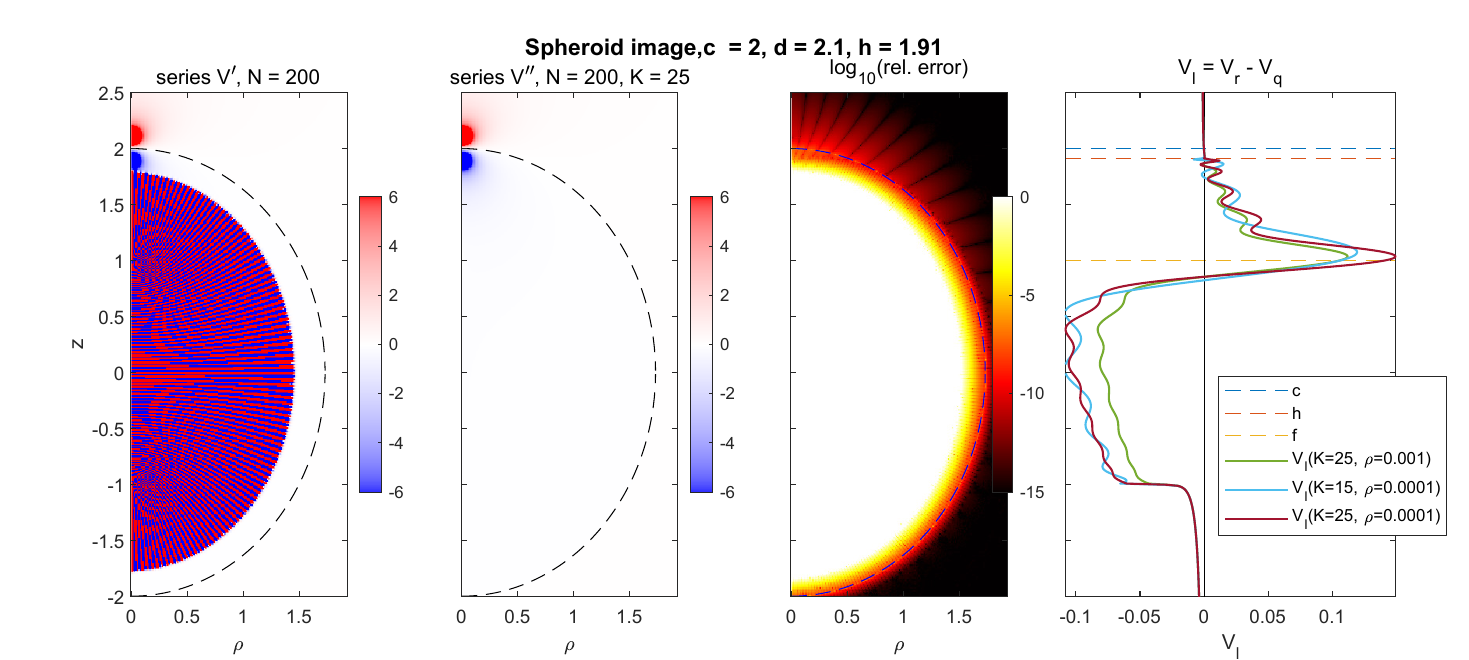}
	\caption{Analytic continuation of $V$ for a point charge near a conducting spheroid, for three aspect ratios and source distances.
		Left-panels: $V$ computed via series \eqref{V'}.
		Left-center panels: series \eqref{V''} showing the analytic continuation of $V$.
		Center-right panels: Relative error between the two series, showing they are equal where they both converge (black regions). 
		%			Center-right panels: the modified series \eqref{V''} is checked against itself, computed with 10 less terms to indicate how well the series has converged - to within about 10\% even very near the singularity.\\
		Far right panels: the potential $V_l$ evaluated very near to the $z$-axis, to get an approximate view of the image line charge density. The series for $V_l$ is truncated at two values of $k$ to show the convergence of the series, and also for two values of $\rho$ to demonstrate that the potential shape doesn't change much, and is therefore evaluated close enough to the axis to consider this a rough estimate of the line charge density.	In all cases $K$ was chosen as high as possible before the catastrophic cancellations became significant.
	} \label{SpheroidImage}
\end{figure}
\thispagestyle{empty}
In order to find the series coefficients, we can use the transformations between spheroidal and spherical harmonics, first transforming the series in to a sum of shifted spherical harmonics, with the following transformation \cite{majic2016thesis,majic2018laplace}:
\begin{align}
Q_n(\xi)P_n(\eta)=\sum_{p=n}^\infty\frac{p!^2}{2(p-n)!(p+n+1)!}\left(\frac{2f}{r'}\right)^{p+1}P_p(\cos\theta') \label{QPvsP_}
\end{align}
where $r'=\sqrt{\rho^2+(z+f)^2}$, $\cos\theta'=(z+f)/r'$.
Then the spherical harmonics can be expanded onto a basis of the stretched spheroidal harmonics \cite{majic2016thesis,majic2017super}:
\begin{align}
\left(\frac{h+f}{r'}\right)^{p+1}P_p(\cos\theta')=\sum_{k=p}^\infty \frac{2(-)^{k+p}(2k+1)(k+p)!}{p!^2(k-p)!}Q_k(\bar\xi)P_k(\bar\eta). \label{PvsQPbar}
\end{align}
Rearranging the summation order, the series \eqref{V'} may be re-expressed as
\begin{align}
V\equiv V''=V_{e}+V^{(0)}-\sum_{k=0}^\infty (2k+1)\sum_{n=0}^k \beta_{kn}(2n+1) \bigg[\frac{Q_n(\xi_d)P_n(\xi_0)}{Q_n(\xi_0)}+q_hP_n(\xi_h)\bigg] ~ Q_k(\bar\xi)P_k(\bar\eta), \label{V''}
\end{align}
where
\begin{align}
\beta_{kn}&=\sum_{p=n}^k \frac{(-)^{p+k}(k+p)!}{(k-p)!(p-n)!(p+n+1)!} x_h^{p+1} \label{beta_kn}\\
\text{and }~x_h&=\frac{2f}{h+f}.\nonumber
\end{align}
These series suffer numerically from catastrophic cancellation, where the individual terms are large but with alternating sign, and their magnitudes are just so that their combined sum is many orders of magnitude smaller than the terms themselves. This cancellation is so dramatic that 15 digits is not enough precision to compute the sum with any accuracy for $k,n\gtrsim20$. Fortunately $\beta_{kn}$ can instead be computed via the following stable recurrence:
\begin{align}
\beta_{k+1,n} &= \frac{k+1}{(n+k+2)(k-n+1)} \bigg\{ (2k+1)\bigg[2x_h - \frac{n(n+1)}{k(k+1)} - 1\bigg]\beta_{kn} 
+ \frac{(n-k+1)(k+n)}{k}\beta_{k-1,n} \bigg\}
\end{align}
with initial values
\begin{align*}
\beta_{nn}=&\frac{1}{2n+1}x_h^{n+1}, \qquad~~
\beta_{n+1,n}=x_h^{n+2}-x_h^{n+1}.
\end{align*}
But another source of numerical instability in \eqref{V''} lies in the sum over $n$, which again suffers from catastrophic cancellation, and this time there seems no way to find an analytic stable method of computation due to the complexity of the terms. The numerical errors become significant for higher $k$, depending on the geometry of the problem.

The line charge distribution is obtained by applying the Havelock formula to \eqref{V''}: 
\begin{align}
\lambda(z)=-\sum_{k=0}^\infty \frac{2k+1}{2}\sum_{n=0}^k \beta_{kn}(2n+1) \bigg[\frac{Q_n(\xi_d)P_n(\xi_0)}{Q_n(\xi_0)}+q_hP_n(\xi_h)\bigg] P_k\left(\frac{2z+f-h}{h+f}\right), \label{varrho2}\qquad -f\leq z<h
\end{align}
which appears numerically to converge as $1/k$ - not fast enough to obtain much detail before numerical problems appear. Even multiple precision has been tried but offers only mild improvement due to it being impractically slow.

Figure \ref{SpheroidImage} shows the analytic continuation of the potential for three different configurations of spheroids and point charges. The nine leftmost panels confirm numerically that $V''$ is the analytic continuation of $V'$ and therefore of $V$. $V''$ is smooth all the way down to the line segment $-f\leq z\leq h$. The far right panels are an attempt to get an idea of what the corresponding image line charge distribution looks like by plotting the potential very close to the axis, using the idea that the potential approaching any line of charge becomes proportional to the line charge density. On these far right panels, the point charges $V_{e}$ and $V^{(0)}$ are subtracted from the total potential $V''$ since they tend to overpower the line source. These plots show that there are no other point sources in the image system unless they are very weak. For all three plotted configurations, the main features of the image line charge distribution are: a positive charge distribution on the segment $f\leq z<h$, which spikes at $z=+f$, then decreases past zero at some $z\lesssim f$ to negative values, and steadily decreases in magnitude as $z\rightarrow-f$. The wave-like oscillations in the right plots are due to the slow convergence of the series. From the right panels we can expect that a simple approximate image charge solution could be found consisting of the negative point charge at $z=h$, plus some positive point charge or dipole at $z=f$ plus a negative uniform line charge $-Q_0(\xi)$, with the exact weighting of each to be optimized numerically. Manual tests for a few source positions and aspect ratios and these image components seem to work very well at satisfying the boundary conditions, but it is beyond the scope of this research to investigate this in detail numerically.

\subsection{Attempt at a first order correction} \label{sec first}
\begin{figure}
	\centering{	\includegraphics[scale=.8]{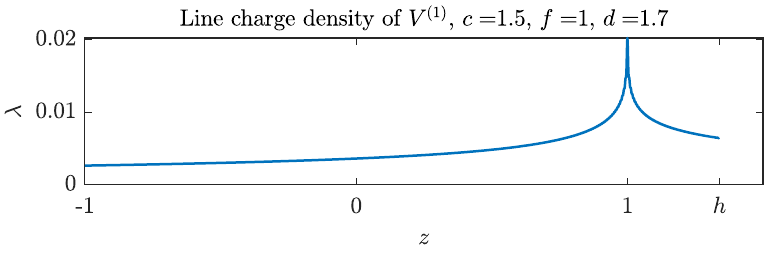}}
	\caption{Line charge density of the first order correction $V^{(1)}$, on the line segment $-f\leq z\leq h$.}\label{fig lamV1}
\end{figure}
Towards the goal of making the series \eqref{varrho2} converge faster so that it may be plotted accurately using double precision, we will look for a first order correction to subtract from the series\footnote{The term ``first order correction" of the series is used loosely here -- it is the correction to the series formed by the series of the first order correction in $n$ to the series terms}. Ideally this is done by analytically determining the next order in the limit of the series coefficients in \eqref{V''}, but this seems intangible. Instead we will follow the approach used to determine $V^{(0)}$, considering the next order correction to the terms in the original series \eqref{Vr}. In that case, subtracting off $V^{(0)}$ also made the ``stretched" series \eqref{V''} converge much faster, since the limit of series coefficients for $V^{(0)}$ expressed in terms of stretched spheroidal harmonics also happen to match the limit of the coefficients for $V$ expressed in terms of stretched spheroidal harmonics. However, this does not happen in the following attempt to extract the next order of $V$. 

From the expansions of the Legendre functions given in \eqref{lim legendre}, we see that the first order correction goes as $1/n$ relative to the leading order, so a simple guess for the first order correction to $V$ would be:
\begin{align}
V^{(1)}=p_h\sum_{n=0}^{\infty}P_n(\xi_h)Q_n(\xi)P_n(\eta) \label{V1prolate}
\end{align}
for some constant $p_h$, which can be determined by matching the limit of the series coefficients of $V^{(1)}$ to those in $V_{r}-V^{(0)}$, to be
\begin{align}
p_h=\frac{q_h}{4}\bigg(\frac{2\xi_0}{\sqrt{\xi_0^2-1}}-\frac{\xi_d}{\sqrt{\xi_0^2-1}}-\frac{\xi_h}{\sqrt{\xi_0^2-1}} \bigg). \label{ph}
\end{align}
Then the series coefficients for $V_{r}-V^{(0)}-V^{(1)}$ converge as $1/n$ relative to the coefficients for $V_{r}-V^{(0)}$. The series \eqref{V1prolate} still diverges inside the spheroid $\xi<\xi_h$ just as does the series for $V_{r}$. We can express \eqref{V1prolate} as a series of stretched spheroidal harmonics (presented without proof):
\begin{align}
V^{(1)}&=p_h x_h\sum_{n=0}^{\infty} P_n(2x_h-1)Q_n(\bar\xi)P_n(\bar\eta) \label{V1bar} 
\end{align}
which converges everywhere except the segment $-f\leq z\leq h$. Its line charge distribution can be computed with no numerical problems and is plotted in Figure \ref{fig lamV1}, showing a spike at $z=f$, which somewhat resembles the spikes in the rightmost panels in Figure \ref{SpheroidImage}.

The problem is that, unlike the case for $V^{(0)}$, the limits of the coefficients of the series' of stretched spheroidal harmonics for $V$ in \eqref{V''} and in \eqref{V1bar} do \textit{not} match.
%, even though their coefficients expressed in the original series \eqref{Vr} and \eqref{V1prolate} \textit{do} match. So unfortunately, subtracting $V^{(1)}$ from $V_{r}-V^{(0)}$ has no effect on the rate of convergence of the stretched series \eqref{V''} or \eqref{varrho2}. 
But there may still exist some first order correction that matches both the first order of the coefficients in \eqref{Vr} and in \eqref{V''}.

\subsection{Point charge off-axis} \label{sec off axis}

For a source off-axis, determining the image is much more complicated. 
We may let the source charge lie at $x_d\geq0,z_d\geq0,y_d=0$, with spheroidal coordinates $\xi_d,\eta_d$. The solution is $V=V_e+V_r$ where
\begin{align}
	V_e=&\frac{f}{\sqrt{(x-x_d)^2+y^2+(z-z_d)^2}}, \nonumber\\
	V_r=&- \sum_{n=0}^\infty\sum_{m=-n}^n (2n+1)(-)^m \frac{(n-m)!^2}{(n+m)!^2}\frac{P_n^m(\xi_0)}{Q_n^m(\xi_0)}Q_n^m(\xi_d)P_n^m(\eta_d) Q_n^m(\xi)P_n^m(\eta)e^{im\phi}. \label{Vr off axis}
\end{align}
In \cite{majic2021imagespheroid}, it was assumed that the domain of convergence could be obtained by analyzing the sum over $n$ for fixed $m$, which lead to a domain of convergence of $\xi>\xi_h$ (Eq. \eqref{xih}), and it was also concluded that for a distant source such that $\xi_d\geq \xi_c=2\xi_0^2-1$, the series converges everywhere except the focal segment. 
However, the true rate of convergence of the sum over $n$ depends on the asymptotics of each sum over $m$ as $n\rightarrow\infty$, which cannot be analyzed easily. The domain of convergence does \textit{not} appear to be $\xi>\xi_h$ from numerical tests, and we will see later this section that it is not the case analytically for long thin spheroids.

%The location of the fundamental singularity/singularities is unknown; a guess is shown in Fig. \ref{fig guess}.
%\begin{figure}[h]
%	\centering{
%		\includegraphics[scale=.35]{ProlateOffAxisClose.pdf}}
%	\caption{A guess of the location of the image of a point charge near a spheroid.} \label{spheroidoffaxis}
%\end{figure} \label{fig guess}

%In comparison to the spheroid with an offset charge (Sec. \ref{ref ellipse outside}), we used a similar approximation to the image point charge like $V_h$ in \eqref{Vh expansion}. While it appears that the image system for the spheroid does contain an image charge of this form, its extraction does not allow analytic continuation of the series, so the image point charge must be attached to a more complex system of image singularities.

While we cannot explicitly determine the domain of convergence of the series \eqref{Vr off axis}, it can still be advantageous to extract an image point charge, similar to \eqref{V^(0)} for the on-axis source, as done in \cite{majic2021imagespheroid}. This image charge is only useful if the source is near the axis or if the spheroid is near spherical.
If we ignore the fact that the series contains terms with unbounded values of $m$, using the asymptotic formulae for the Legendre functions for large $n$ with $m$ fixed \eqref{lim legendre}, we can get an approximation for the individual terms as $n\rightarrow\infty$:
\begin{align}
	V_r\approx V^{(0)}&=q_h\sum_{n=0}^\infty\sum_{m=-n}^n (2n+1)(-)^m \frac{(n-m)!^2}{(n+m)!^2}P_n^m(\xi_h)P_n^m(\eta_d) Q_n^m(\xi)P_n^m(\eta)e^{im\phi}\\
	&=\frac{q_hf}{\sqrt{(x-x_h)^2+y^2+(z-z_h)^2}} \label{V0 off axis}
\end{align}
where again
$\xi_h=(2\xi_0^2-1)\xi_d - 2\xi_0\sqrt{\xi_0^2-1}\sqrt{\xi_d^2-1}$, $~q_h=-((\xi_h^2-1)/(\xi_d^2-1))^{1/4}$, and
$x_h=f\sqrt{\xi_h^2-1}\sqrt{1-\eta_d^2}$, $z_h=f\xi_h\eta_d$. This proposed image point charge lies on the same coordinate $\eta=\eta_d$ as the source charge. 
As done in \cite{lindell2001electrostatic} for the axial case, the approximation \eqref{V0 off axis} can be subtracted from the series \eqref{Vr off axis} to give
\begin{align}
 	V_r=&\frac{q_hf}{\sqrt{(x-x_h)^2+y^2+(z-z_h)^2}} \nonumber\\
 	& - \sum_{n=0}^\infty\sum_{m=-n}^n (2n+1)(-)^m \frac{(n-m)!^2}{(n+m)!^2}\left(\frac{P_n^m(\xi_0)}{Q_n^m(\xi_0)}Q_n^m(\xi_d)+q_hP_n^m(\xi_h)\right)P_n^m(\eta_d) Q_n^m(\xi)P_n^m(\eta)e^{im\phi}. \label{Vrp off axis}
\end{align}
However, this only gives a faster converging series over $n$ for each $m$ individually and does not necessarily make the double series itself converge faster, unless the source is near the axis or the spheroid is near spherical. 
%And again since we do not know the exact domain of convergence, $x=x_h,z=z_h$ is only an approximate location of the image

For a thin spheroid with $a<<c$, we will consider for simplicity a relatively close point source on the $x$ axis at $a<x_d<<c$. Here we have $\xi,\xi_d\rightarrow1$, $\eta_d=0$, and we look at $\eta=0$ to further simplify the problem.
In this limit the Legendre functions behave as \cite{NIST:DLMF}
\begin{align}
	P_n^m(\xi)&\rightarrow\frac{(n+m)!}{(n-m)!m!}\left(\frac{\xi-1}{2}\right)^{m/2} \qquad \xi\rightarrow1\\
	Q_n^m(\xi)&\rightarrow\frac{(m-1)!}{2}\left(\frac{\xi-1}{2}\right)^{-m/2}  \qquad \xi\rightarrow1 \\
\text{and } ~~	P_n^m(0)&=(-)^\frac{n-m}{2}\frac{(n+m-1)!!}{(n-m)!!}
\end{align}

This leads to a limiting form of the coefficients in the series of 
\begin{align}
	V_r\rightarrow -\sum_{n=0}^\infty \sum_{m=0}^n (2-\delta_{m0}) \frac{2n+1}{2m}(-)^m \left(\frac{(\xi_0-1)^2}{(\xi-1)(\xi_d-1)}\right)^{m/2}P_n^m(0)P_n^{-m}(0) \label{Vroffaxislim}
\end{align}
We cannot state that the series converge without further analysis of the double summation, but
a simple observation is that the series over $m$ diverges at least if $(\xi_0-1)^2/[(\xi-1)(\xi_d-1)]>1$, 
or if
\begin{align}
	\xi<\xi_i=\sqrt{\frac{(\xi_0^2-1)^2}{\xi_d^2-1}+1} \label{xi_i}
\end{align}
(which is an equivalent condition as $\xi_0,\xi_d\rightarrow1$, and appears slightly more accurate for larger $\xi_0,\xi_d$).
This is satisfiable for all $\xi_0,\xi_d$, i.e. the series will always diverge for some region near the focus. Note that this only analysis only applies to thin spheroids where the source is off axis and close to the surface.
This disproves the convergence condition $\xi>\xi_h$; for example if $\xi_d=2\xi_0^2-1$ then $\xi_h=1$ but the solution does not converge for all $\xi>1$. 
The condition for divergence $\xi<\xi_i$ is reasonable for thin spheroids, and is consistent with the condition for divergence of the solution for a point charge near an infinite cylinder. That is, if we express the $\xi$ coordinates in terms of their $x$ values on the $z=0$ plane, we can conclude that the solution diverges if $x>a^2/x_d$, where $x_d$ is the $x$ coordinate of the source. This is also the condition for divergence of the solution for point charge at $x=x_d$ outside a cylinder of radius $a$ \cite{majic2020cylinder}.

\begin{figure}
	\begin{minipage}{.7\textwidth}
		\includegraphics[scale=.73]{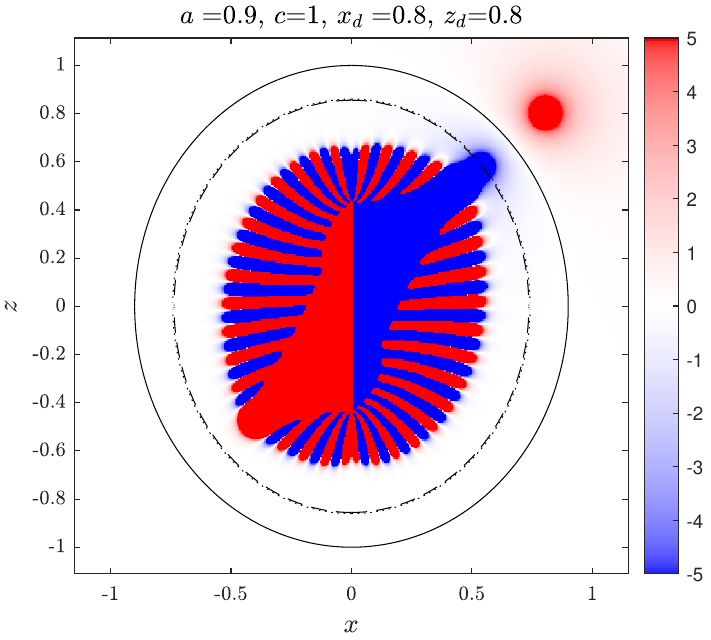}\\
		\includegraphics[scale=.73]{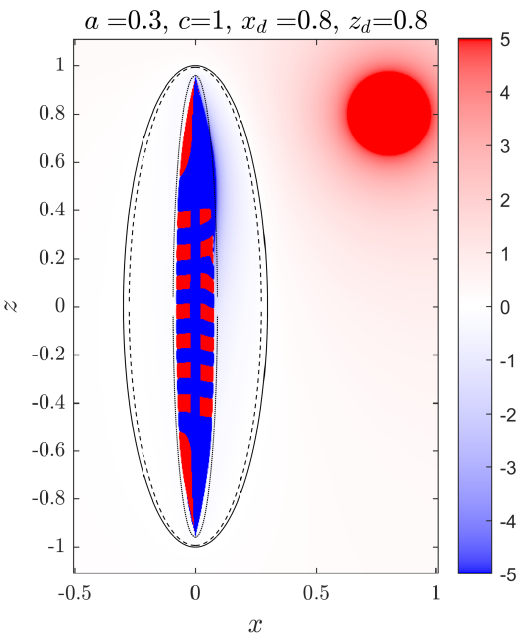}
	\end{minipage}
	\begin{minipage}{.29\textwidth}
		\includegraphics[scale=.68]{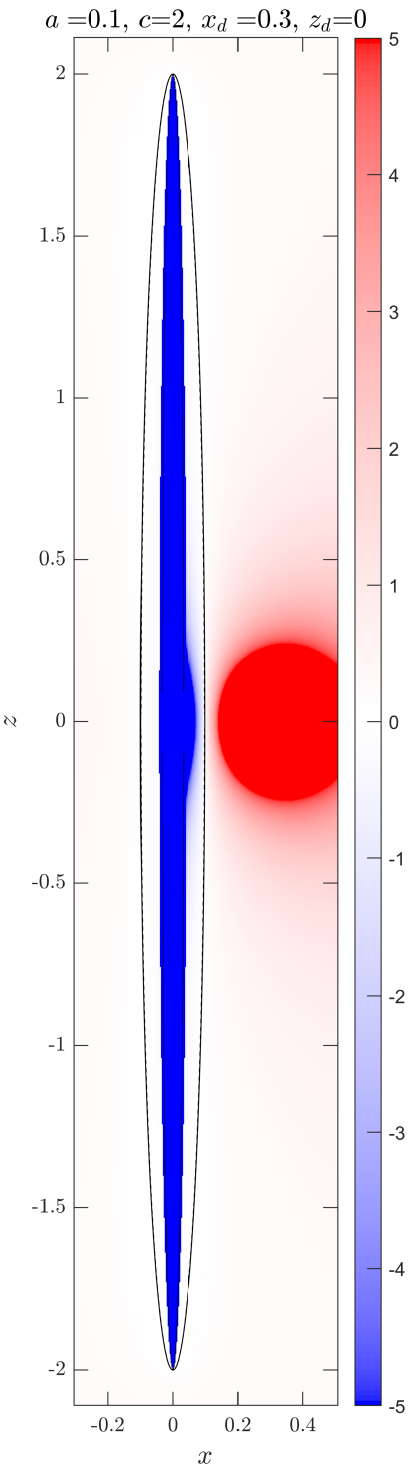}
	\end{minipage}
	\caption{Exact potential $V$ for a point charge outside a conducting prolate spheroid, computed via \eqref{Vr off axis} with $n,m\leq 85$ for the higher two aspect ratios, or via \eqref{Vrp off axis} with $n,m\leq35$ for the lowest aspect ratio. In each plot the solid black line is the spheroid boundary $\xi_0$, the dashed line is $\xi_h$ and the dotted line is $\xi_i$.} \label{fig V off axis}
\end{figure}

The solution \eqref{Vr off axis} is plotted in Figure \ref{fig V off axis} for three different aspect ratios. For the low aspect ratios we partially see something like an image point charge at $x\approx x_h,z \approx z_h$.
For higher aspect ratios, we see that the solution does converge down to $x=a^2/x_d$ as we reasoned is possible for thin spheroids, and certainly one can conclude that the condition for convergence is not $\xi>\xi_h$. 
For low aspect ratios, $\xi_i\rightarrow\xi_h$, which both appear to lie on the extent of the domain of convergence, while for higher aspect ratios, $\xi_h$ is way off and $\xi_i$ is reasonably close. It also appears that convergence could depend on $\eta$ and $\phi$, which is possible analytically if $\eta$ and $\phi$ were to change the behavior of the sum over $m$ as $n\rightarrow\infty$.

Only up to $n,m=85$ could be used in Figure \ref{fig V off axis} because the factor $(n+m)!$ overflows for any higher $n,m$. This problem could be avoided by using normalized Legendre functions \cite{fukushima2014prolate}, which would allow more terms to be included in the series, giving a clearer picture of the boundary between convergence and divergence.

\section{Point charge inside spheroid} \label{sec inside}
We now move on to sources placed inside a conducting spheroidal boundary, where the image systems lie outside the spheroid. 
\subsection{Point charge at the center} \label{sec center}
\begin{figure}
	\centering{\includegraphics[scale=0.52]{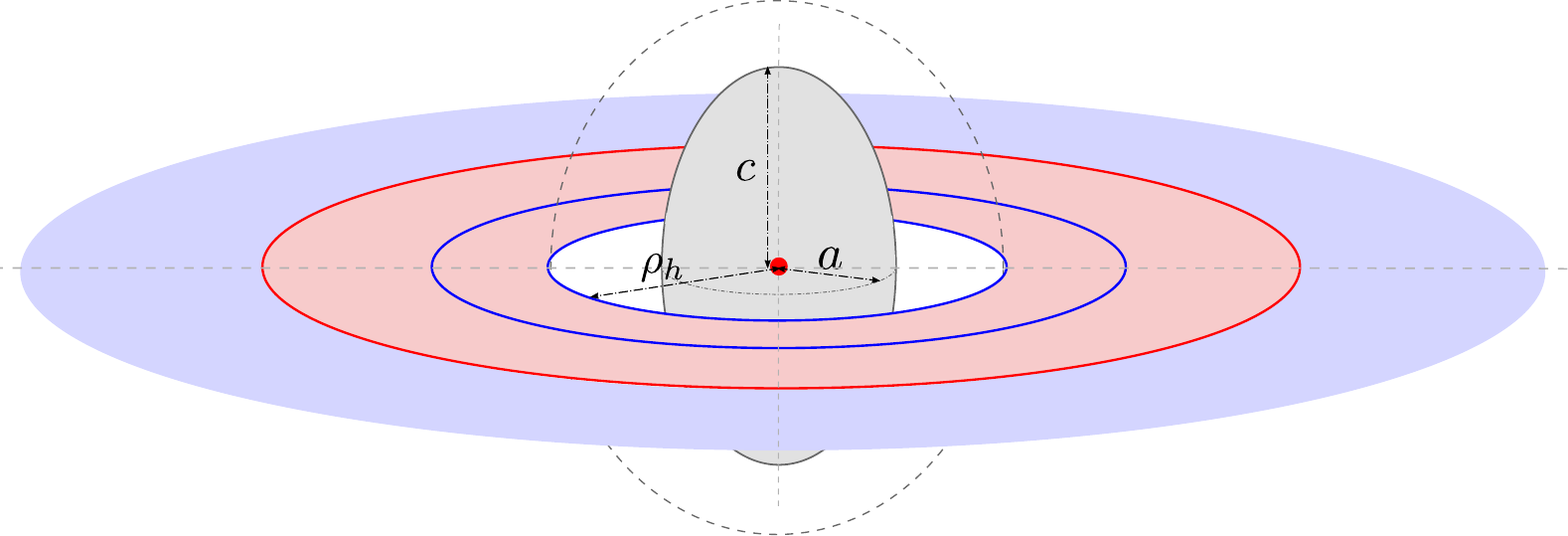}}
	\caption{Schematic of the problem of a point charge centered in a prolate spheroid, roughly showing the proposed disk image with singular rings that extends out to infinity. Red represents positive charge and blue negative.}\label{SpheroidDiskImage}
\end{figure}

The simplest case is that of a point charge located exactly at the center. The excitation potential is
\begin{align}
V_{e}=\frac{f}{r} =\sum_{n=0}^\infty (2n+1) P_n(0) Q_n(\xi)P_n(\eta).  \qquad \label{Ve center}
\end{align}
Note that for $n$ odd, $P_n(0)=0$.
The total potential inside is $V=V_{e}+V_{r}$, where
\begin{align}
V_{r}=-\sum_{n=0}^\infty (2n+1)\frac{Q_n(\xi_0)}{P_n(\xi_0)}P_n(0)P_n(\xi)P_n(\eta), \label{Vr center}
\end{align}
which converges inside a spheroidal volume with surface $\xi_h=2\xi_0^2-1$. Now we want to determine the form of the image, which must lie outside this spheroid $\xi=\xi_h$. For the similar problem of a charge located inside a cylinder \cite{majic2020cylinder}, the image lies on the $z=0$ plane with a circular hole. So at least for thin long spheroids we should expect a similar disk image, where the hole touches the spheroid of divergence. So we will assume that the image lies on
\begin{align}
\rho\geq \rho_h=f\sqrt{\xi_h^2-1}, \quad z=0, \label{singular disk domain}
\end{align}
which is depicted in Figure \ref{SpheroidDiskImage}. But before deriving the exact analytic continuation of $V_r$, lets consider some simple approximations.

\subsection{leading order approximation}
Here we look for an approximation analogous to that of the point charge approximation for the source outside, by analyzing the limit of the series coefficients as $n\rightarrow\infty$.
%We can improve the numerical stability of \eqref{Vr center} slightly by considering a simple approximation to $V_{r}$, similar to both the image charge $V^{(0)}$ in Sec. \ref{sec lindell} and 
In addition to the limits for the Legendre functions in \eqref{lim legendre}, we need the limit for $P_n(0)$:
\begin{align}
	P_n(0)\rightarrow&~\sqrt{\frac{2}{n\pi}}(-)^{n/2} \left[1-\frac{1}{4n}+\cO(n^{-2})\right]. \label{lim P0}
\end{align}
Then as $n\rightarrow\infty$ the coefficients in the series \eqref{Vr center} go to leading order as:
\begin{align}
	\frac{Q_n(\xi_0)}{P_n(\xi_0)}P_n(0) \rightarrow \frac{\sqrt{2\pi/n}(-)^{n/2}}{(\xi_0+\sqrt{\xi_0^2-1})^{2n+1}}. \label{limitVr center}
\end{align}
We want to find a simple function that matches this limit, that when substituted in \eqref{Vr center}, has a simple closed form for the corresponding series. 
For the infinite conducting cylinder, the leading order approximation is two point charges offset on the imaginary $z$-axis at $z=\pm 2i$ \cite{majic2020cylinder}, which appears in real space as an excavated disk. So for the spheroid, we find that two point charges offset by $z=\pm i\rho_h$ is a good approximation. 
The spheroidal series expansion for this is easily found from the inverse distance expansion \cite{Morse1953}:  
%\begin{align}
%	\frac{c}{\sqrt{\rho^2+(z\pm id)^2}}=\sum_{n=0}^{\infty}(\pm)^n(2n+1)Q_n(i d)P_n(\xi)P_n(\eta) \label{GF imag prolate}
%\end{align}
\begin{align}
	V^{(0)}&=\frac{if\sqrt{\xi_h}}{\sqrt{\rho^2+(z-i\rho_h)^2}} - \frac{if\sqrt{\xi_h}}{\sqrt{\rho^2+(z+i\rho_h)^2}} \nonumber\\
	&=-2i\sqrt{\xi_h} \sum_{n=0}^{\infty}(2n+1)Q_n(i \rho_h/f)P_n(\xi)P_n(\eta), \label{V0center}
\end{align}
where we have introduced prefactors so the series coefficients match those in \eqref{limitVr center} in the limit $n\rightarrow\infty$.
%Although not immediately apparent, the series coefficients in \eqref{V0center} have the same asymptotic limit as \eqref{limitVr center}, since $(\rho_h/f+\sqrt{\rho_h^2/f^2+1})^{n+1/2}=(\xi_0+\sqrt{\xi_0^2-1})^{2n+1}$. 
$V^{(0)}$ is a good approximation to $V_{r}$ for fatter spheroids, accurate to within 1\% 
%as shown in Figure \ref{V0 plots}, 
(a constant correction term should be applied to $V^{(0)}$, replacing the $n=0$ term in \eqref{V0center} with the term in \eqref{Vr center}). The approximation works better for rounder spheroids; for $c=2f$, the approximation is visually indistinguishable from the exact analytic continuation plotted in Figure \ref{fig Vcenter}.
%This approximation is similar to that used for the image of a point charge on the axis of an infinite cylinder \cite{majic2020cylinder}, which makes sense because both problems are axially symmetric and the cylinder can be seen as the limit of a ling spheroid. 
In fact \eqref{V0center} reduces to the image approximation for the cylinder if we take $c\rightarrow\infty$ with $a$ fixed.

\subsection{First order correction}
A first order correction can be found in a similar manner, by considering the next order in the coefficients for the Legendre functions, which allows us to deduce the limit of the coefficients in $V_r-V^{(0)}$:
\begin{align}
	(2n+1)\bigg[\frac{Q_n(\xi_0)}{P_n(\xi_0)}P_n(0)- 2i\sqrt{\xi_h}Q_n(i\rho_h/f)\bigg] \rightarrow ~
	\sqrt{\frac{\pi}{2n}}(-)^{n/2}\frac{\xi_0^3}{\xi_h^{3/2}}\frac{\Big(\xi_0+\sqrt{\xi_0^2-1}\Big)^{-2n-1}}{\sqrt{\xi_0^2-1}}. \label{lim O1}
\end{align}
Now consider the limit of
\begin{align}
	(2n+1)\int Q_n(u)\d u ~=~Q_{n+1}(u)-Q_{n-1}(u)~ \rightarrow~ \sqrt{\frac{2\pi}{n}}\frac{(u^2-1)^{1/4}}{(u+\sqrt{u^2-1})^{n+1/2}} ,
\end{align}
which can be matched to the limit of \eqref{lim O1} with $u=i\rho_h/f$ and the right prefactor. This leads to the first order correction $V^{(1)}$, writing $V_{r}\approx V^{(0)}+V^{(1)}$:
\begin{align}
	V^{(1)} = -\frac{\xi_0^3}{2\xi_h^{3/2}\sqrt{\xi_0^2-1}}\sum_{n=0}^\infty  (2n+1)\int^{i\rho_h/f} Q_n(u)\d u ~ P_n(\xi)P_n(\eta). \label{V1series}
\end{align}
(For $n=0$, $\int^{i\rho_h/f} Q_0(x) \d x = 1 - \rho_h/f\tan^{-1}(f/\rho_h) - \log(1+f^2/\rho_h^2)/2 $). 
The series for $V^{(1)}$ has a simple closed form expression:
\begin{align}
	V^{(1)}=\frac{\xi_0^3}{2\xi_h^{3/2}\sqrt{\xi_0^2-1}}\text{Re}\big\{\log\frac{2\rho_h}{\rho_h-iz+\sqrt{\rho_h^2-2iz\rho_h-r^2}} \big\} \label{V1},
\end{align}
which is also singular on the inverted disk $\rho>\rho_h$, $z=0$. \eqref{V1} can be verified by expanding \eqref{V1series} onto spherical harmonics and comparing with the first order correction for the cylinder in \cite{majic2020cylinder}. $V^{(1)}$ has a positive surface charge density that decreases with $\rho$.
%Adding $V^{(1)}$ to $V^{(0)}$ increases the accuracy of the approximation.
%, and subtracting $V^{(1)}$ from $V_r$ makes the original series \eqref{Vr} converge faster (after already subtracting off $V^{(0)}$). 

\subsection{Analytic continuation - image disk}
For the 2D problem for a source centered in an ellipse, we discovered that the image system was a series of point charges on the $y$-axis - the thin axis of the ellipse. An intuitive generalization of this to the prolate spheroid would be a series of rings on the $xy$-plane over the thin axes of the spheroid. (For an oblate spheroid we might expect the images to lie on the $z$-axis but that is outside our scope).

To find a mathematical expression for the analytic continuation of $V_{r}$, we could follow the approach for the infinite cylinder \cite{majic2020cylinder} and search for an expression as a sum or integral of cylindrical harmonics $J_0(k \rho)e^{-k |z|}$, since this expression should converge everywhere except this image disk. However, there isn't a clear ansatz for this is -- for the cylinder the solution is a discrete sum over $k=k_n$, the $n^{th}$ zeros of $J_0$, but there seems no reason why $V_{r}$ couldn't be a sum over different values of $k$, or even an integral over $0\leq k<\infty$. 
So instead we will use a different basis that also shares this plane singularity - radially inverted irregular oblate spheroidal harmonics (RIIOSHs). They are singular on the $z=0$ plane and also come with a hole in the middle, which can be fitted to match the assumed image domain \eqref{singular disk domain} exactly. The RIIOSHs are $r^{-1}Q_n(i\bar{\chi})P_n(\bar{\zeta})$, where $\hat{\chi}, \hat{\zeta}$ are radially inverted oblate spheroidal coordinates:
\begin{align}
\hat{\chi }&=\frac{1}{\sqrt{2}\rho_h}\sqrt{\hat{r}^2-\rho_h^2+\sqrt{(\hat{r}^2-\rho_h^2)^2+(2\rho_h u\hat{r})^2}}, \qquad
\hat{\zeta}=\frac{u \hat{r}}{\rho_h\hat{\chi}}, \qquad
\hat{r}=\frac{\rho_h^2}{r}.
\end{align}
$\rho_h$ is the focal disk radius which has been set so that the RIIOSHs are then singular on the domain \eqref{singular disk domain}.
To expand $V_{r}$ in terms of RIIOSHs, we first expand the prolate spheroidal harmonics in \eqref{Vr center} in terms of regular spherical harmonics \cite{buchdahl1977relation}, and
%Inserting this expansion and rearranging the summation order gives
%\begin{align}
%	V_{r}=-\sum_{p=0:2}^\infty \Bigg[ \sum_{n=p:2}^\infty (2n+1)\frac{Q_n(\xi_0)}{P_n(\xi_0)}P_n(0) \frac{(-)^\frac{n-p}{2} (n+p-1)!!}{p!(n-p)!!} \Bigg]\left(\frac{r}{f}\right)^p P_p(\cos\theta) \label{Vspherical}
%\end{align}
%This spherical series should converge inside the sphere $\rho<d$, which is a smaller domain than the original prolate spheroidal series \eqref{Vr center}, not offering any analytic continuation of $V_{r}$. 
then expand the spherical harmonics in terms of RIIOSHs. The required expansion can be obtained by applying radial inversion to a well known transformation formula between the \textit{irregular} spherical harmonics and the irregular (external) oblate spheroidal harmonics \cite{jeffery1916relations}. Inserting these expansions into \eqref{Vr center} and rearranging the summation order gives

\begin{align}
V_{r}=\sum_{k=0:2}^\infty \Bigg\{\sum_{p=0:2}^k  \Bigg[\sum_{n=p:2}^\infty (2n+1)\frac{Q_n(\xi_0)}{P_n(\xi_0)}&P_n(0) \frac{(-)^\frac{n-p}{2} (n+p-1)!!}{p!(n-p)!!}\Bigg] \nonumber\\
&\times\frac{(k+p-1)!!}{p!(k-p)!!}\left(\frac{\rho_h}{f}\right)^k  \Bigg\} ~ (2k+1)i^{k+1}~\frac{\rho_h}{r}Q_k(i\hat{\chi})P_k(\hat{\zeta}). \label{Voblate}
\end{align}
The series over $n$ appears to converge quickly and be numerically stable (checked for $p\leq100$). But the sum over $p$ suffers from catastrophic cancellation like in \eqref{V''} - the terms are only accurate for $k\lesssim 44$. Still, \eqref{Voblate} offers a significant analytic continuation of the original series \eqref{Vr center}.
We can also improve the rate of convergence of \eqref{Voblate} by adding and subtracting the two forms for $V^{(0)}$  in \eqref{V0center}:
\begin{align}
V_{r}=\frac{ic\sqrt{\xi_h}}{\sqrt{\rho^2+(z-i\rho_h)^2}}
&- \frac{ic\sqrt{\xi_h}}{\sqrt{\rho^2+(z+i\rho_h)^2}} \nonumber\\-\sum_{k=0:2}^\infty \Bigg\{&\sum_{p=0:2}^k  \Bigg[\sum_{n=p:2}^\infty (2n+1)\bigg(\frac{Q_n(\xi_0)}{P_n(\xi_0)}P_n(0)-2i\sqrt{\xi_h}Q_n(i\rho_h/f) \bigg)\nonumber\\
&\times \frac{(-)^\frac{n-p}{2} (n+p-1)!!}{p!(n-p)!!} \Bigg] \frac{(k+p-1)!!}{p!(k-p)!!}\left(\frac{\rho_h}{f}\right)^p  \Bigg\}~ (2k+1)i^{k+1}~\frac{\rho_h}{r}Q_k(i\hat{\chi})P_k(\hat{\zeta}). \label{VoblateV0}
\end{align}
%\begin{align}
%V_{r}=&V^{(0)}+V^{(1)}-\sum_{k=0:2}^\infty \Bigg\{\sum_{p=0:2}^k  \Bigg[\sum_{n=p:2}^\infty (2n+1)\bigg(\frac{Q_n(\xi_0)}{P_n(\xi_0)}P_n(0)-2i\sqrt{\xi_h}Q_n(ih/f)+\frac{\xi_0^3\int^{ih/f} Q_n(x)\d x}{2\xi_h^{3/2}\sqrt{\xi_0^2-1}} \bigg)\nonumber\\
%&\times \frac{(-)^\frac{n-p}{2} (n+p-1)!!}{p!(n-p)!!} \Bigg] \frac{(k+p-1)!!}{p!(k-p)!!}\left(\frac{h}{f}\right)^p  \Bigg\}~ (2k+1)i^{k+1}~\frac{h}{r}Q_k(i\hat{\chi})P_k(\hat{\zeta}). \label{VoblateV1} % worse than VoblateV0
%\end{align}
\begin{figure}
		\includegraphics[scale=1,trim={0 .45cm 0 0}, clip]{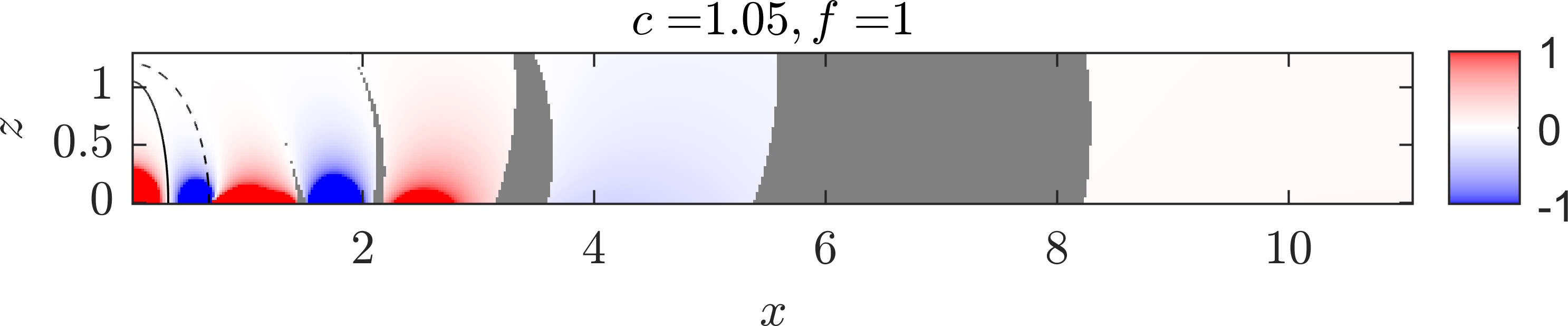}
		\includegraphics[scale=1,trim={0 .45cm 0 0}, clip]{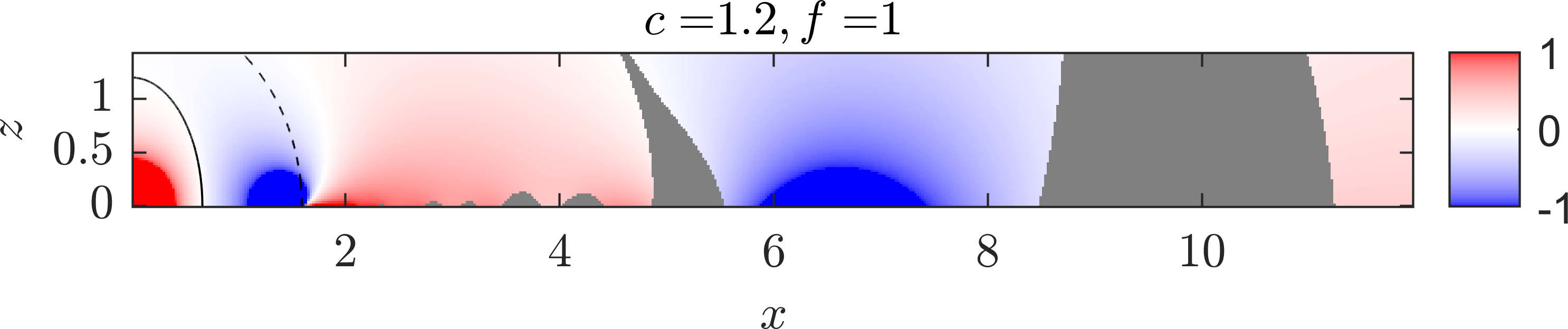}\\
		\includegraphics[scale=1,trim={0 0 0 0}, clip]{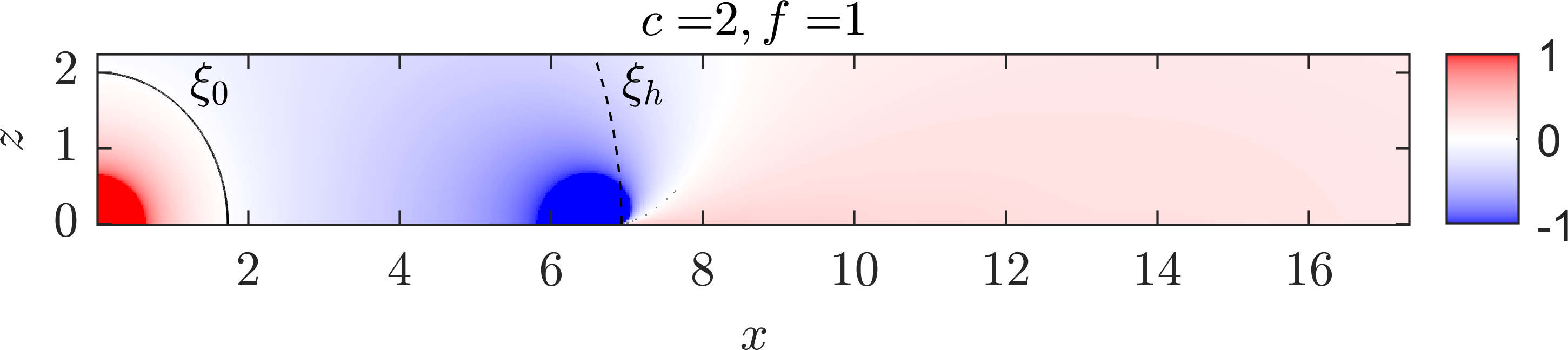}
	\caption{The analytic continuation of $V$ using Eq. \eqref{VoblateV0} for spheroids of different aspect ratios. The series \eqref{VoblateV0} is computed up to $k=K$ for two similar values of  $K$ (top: $K=34,38$, middle $K=40,44$, bottom: $K=36,40$) and these are compared to check that the series has converged sufficiently and has not encountered numerical cancellations. Regions where the two series have converged to less than 10\% of each other have been grayed-out to avoid showing incorrect data. There is no guarantee that the remaining plot is accurate to within 10\% error, since the two series may just happen to coincide, likely in the areas between the gray lobes.}\label{fig Vcenter}
\end{figure}
The faster convergence means \eqref{VoblateV0} can accurately compute $V_{r}$ in a larger domain relative to \eqref{Voblate} before numerical cancellations become a problem.
Eq. \eqref{VoblateV0} is plotted in Figure \ref{fig Vcenter} for spheroids of different aspect ratios. Numerical errors are high where $\hat\chi$ is small, near the image disk, 
%but also as $r\rightarrow\infty$ in all directions. 
%This is because the terms in the series with higher $k$ are more important, and these terms are where numerical cancellations in the series coefficients are worst.  
%So unfortunately the RIIOSH series fails near the proposed image disk, 
but we can still make loose deductions of image charge density from the plots. $V_{r}$ changes sign as it moves out in the $\rho$ direction, indicating that the image disk charges also change sign in this way. A similar pattern is seen for the infinite cylinder, but there the image changes sign at regularly spaced intervals, while the pattern here appears to stretch as $\rho$ increases. In the cylinder the surface charge density also became infinite at these evenly spaced intervals to make singular rings, so we might expect these to occur for the spheroid (and there is at least one singular ring at $\xi_h$). From the 2D problem of a source centered in an ellipse, we could guess that the image rings would lie at $\text{acosh}\xi=2k\text{acosh}\xi_0$ (and $\eta=0$) -- at least the first image ring at $\text{acosh}\xi_h=2\text{acosh}\xi_0$ agrees with this hypothesis.

Unfortunately like in Section \ref{sec first}, subtracting $V^{(1)}$ does not make the RIIOSH series converge faster, although $V^{(0)}+V^{(1)}$ does match the analytic continuation well for the inner part of the image disk.

In order to prove that the image lies completely on the $z=0$ plane, we would have to show that the coefficients in either series \eqref{Voblate} or \eqref{VoblateV0} grow at less than an exponential rate as $k\rightarrow\infty$ so that the $Q_k(i\hat\chi)$ will make the series converge. This does appear to be true numerically for the values of $k$ we can compute without catastrophic cancellation. 

%For a point charge located off-center in the prolate spheroid at some $\eta_d$, but still on the rotation axis, we can only guess where the image lies. Spatial plots of the standard series solutions reveal just the edge of the image, and it appears that this edge lies on the same coordinate line $\eta=\eta_d$. Perhaps the image lies on the surface $\eta=\eta_d$, $\xi>\xi_h$ - an infinite parabolic curved sheet with a circular hole around the spheroid. For $\eta_d=0$, this reduces to the flat disk image proposed in this section, and for a point charge located at the focus $\eta=1$, reduces to a straight line extending up the $z$ axis for $\xi>\xi_h$.
%\FloatBarrier

\subsection{Point charge off axis inside spheroid} \label{sec inside off axis}
\begin{figure}
	\includegraphics[scale=.66]{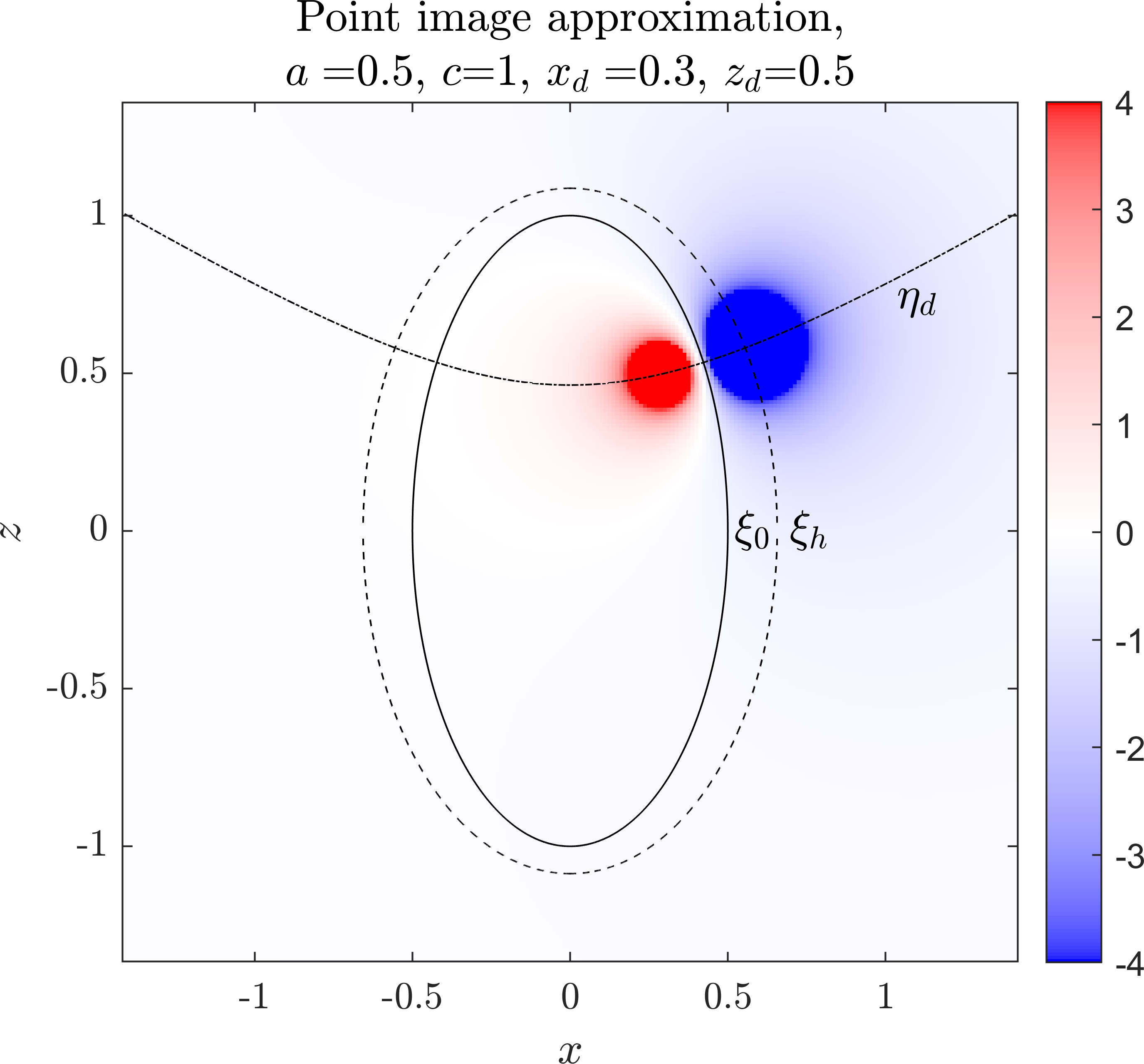}
	\includegraphics[scale=.66]{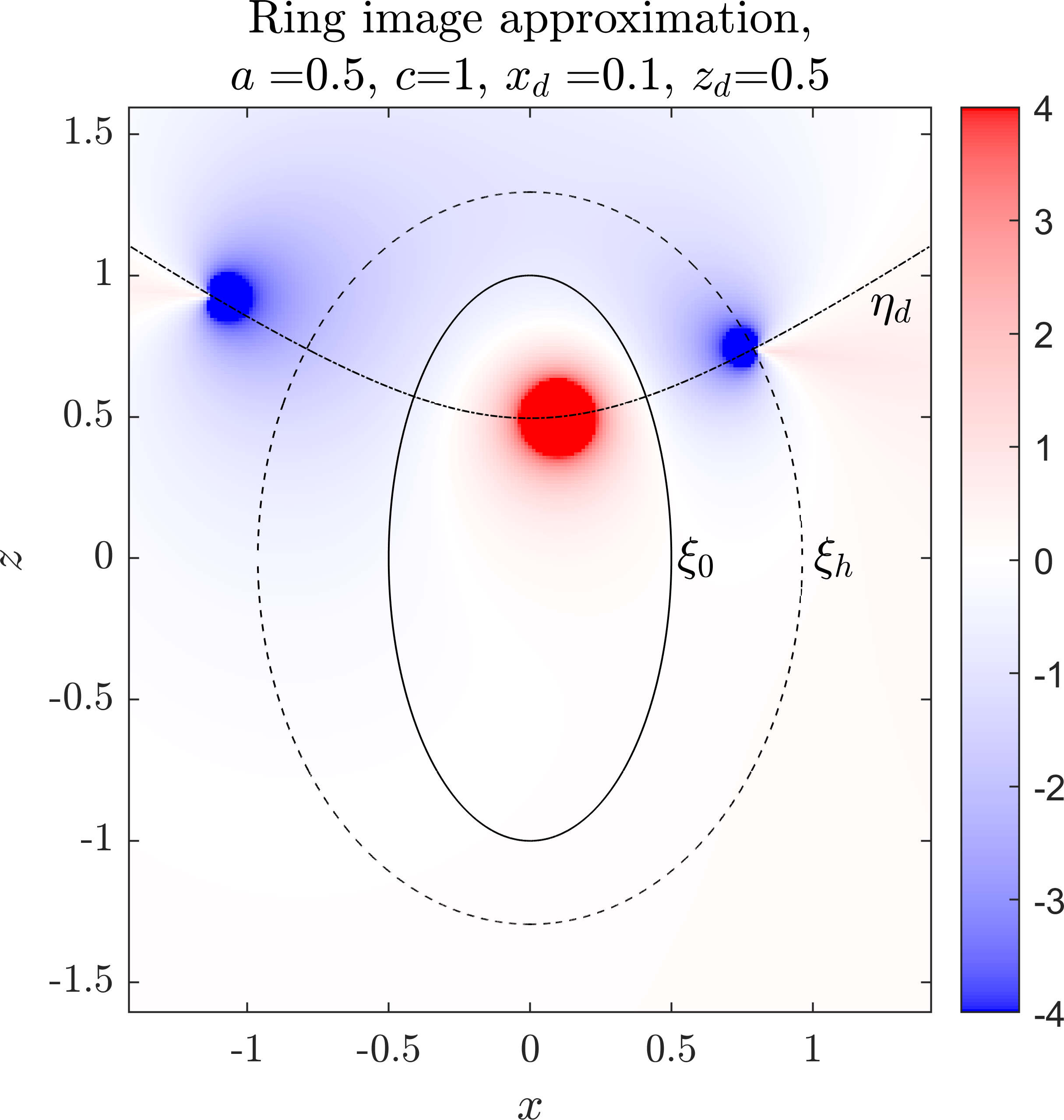}
	\caption{Color plots of the image approximations to the potential of a point charge inside a conducting spheroid. Left: the real point charge approximation $\dot V_r^0+V_e$ using equation \eqref{Vr0 inside} for a point charge near the surface. Right: the complex point charge or ring approximation $\mathring V_r^0+V_e$ using equation \eqref{Vr0b inside} for a point charge nearer the focus.} \label{fig image real complex}
\end{figure}
\FloatBarrier

We can extend some of the concepts of the previous sections to offset charges. For the ellipse, the images were point sources on a hyperbola, so we might expect that the images for the spheroid would lie on the hyperboloid $\eta=\eta_d$. But in this section we only manage to find an approximate image ring.
Let the source charge lie at $x=x_d>0,~y=0,~z=z_d>0$, or equivalently at $\eta=\eta_d>0,~\xi=\xi_d,~\phi=0$. The excitation potential is
\begin{align}
	V_e=\frac{f}{\sqrt{(x-x_d)^2+y^2+(z-z_d)^2}},
\end{align}
and the reflected potential is
\begin{align}
	V_r=-\sum_{n=0}^\infty\sum_{m=-n}^n (-)^m \frac{(n-m)!^2}{(n+m)!^2} \frac{Q_n^m(\xi_0)}{P_n^m(\xi_0)}P_n^m(\xi_d)P_n^m(\eta_d) P_n^m(\xi)P_n^m(\eta)\cos m\phi. \label{Vr inside}
\end{align}
Like for the off-axis external source in Section \ref{sec off axis}, the exact rate of convergence depends on the values of the whole sum over $m$ as $n\rightarrow\infty$, but this is not straightforward to determine analytically. Instead we will use a crude analysis by considering the limit as $n\rightarrow\infty$ for single terms with fixed $m$. In this case the series over $n$ diverges outside the external spheroid $\xi_h=(2\xi_0^2-1)\xi_d - 2\xi_0\sqrt{\xi_0^2-1}\sqrt{\xi_d^2-1}$, or $\acosh\xi_h=2\acosh\xi_0-\acosh\xi_d$.
In contrast to the case of an external off axis source, $\xi_h$ appears to closely fit the boundary of divergence of the series \eqref{Vr inside}, from numerical tests across a wide range of aspect ratios and source positions.

Next we investigate two ways to approximate the terms in this sum as $n\rightarrow\infty$ for $m$ fixed, which work better for different locations of the source. 
\subsubsection{Point charge near surface}
When the source charge is near the surface, we use an approximation of a point charge somewhat opposite the source charge.
In the series \eqref{Vr inside}, we replace the coefficients with
\begin{align}
	-\frac{Q_n^m(\xi_0)}{P_n^m(\xi_0)}P_n^m(\xi_d)\rightarrow q_h Q_n^m(\xi_h)\big[1+\cO(1/n)\big], \label{Vr0 in lim}
\end{align}
where $q_h=-((\xi_h^2-1)/(\xi_d^2-1))^{1/4}$ is the strength of the image charge.
Substituting this approximation into \eqref{Vr inside} gives an approximate image potential $\dot V_r^0$:
\begin{align}
	\dot V_r^0&=q_h\sum_{n=0}^\infty\sum_{m=-n}^n \frac{(n-m)!^2}{(n+m)!^2}(-)^m Q_n^m(\xi_h)P_n^m(\eta_d) P_n^m(\xi)P_n^m(\eta)\cos m\phi \nonumber\\
	&=\frac{q_hf}{\sqrt{(x-x_h)^2+y^2+(z-z_h)^2}}\label{Vr0 inside}
\end{align}
where $x_h=f\sqrt{\xi_h^2-1}\sqrt{1-\eta_d^2},~z_h=f\xi_h\eta_d$. We can then express the exact potential $V_r$ as a faster converging series by adding and subtracting the different forms of $\dot V_r^0$:
\begin{align}
	V_r=&\frac{fq_h}{\sqrt{(x-x_h)^2+y^2+(z-z_h)^2}}\nonumber\\
	&-\sum_{n=0}^\infty\sum_{m=-n}^n \frac{(n-m)!^2}{(n+m)!^2}(-)^m(2n+1) \bigg(\frac{Q_n^m(\xi_0)}{P_n^m(\xi_0)}P_n^m(\xi_d)+q_hQ_n^m(\xi_h)\bigg)P_n^m(\eta_d) P_n^m(\xi)P_n^m(\eta)\cos m\phi. \label{Vrp in}
\end{align}

 But \eqref{Vr0 inside} fails when the source charge is near the focus of the spheroid, since the problem becomes more rotationally symmetric, and a single image point charge placed off axis will break the symmetry. 
 
\subsubsection{Point charge near focus}
 In this case we look to an approximation that is a generalization of the ring image \eqref{V0center} for the source at the center. 
The coefficients in \eqref{Vr inside} can instead be approximated as
\begin{align}
	-\frac{Q_n^m(\xi_0)}{P_n^m(\xi_0)}P_n^m(\eta_d)\rightarrow q_r Q_n^m(\xi_r)\big[1+\cO(1/n)\big], \label{Vr0b app}
\end{align}
where we use new coordinates with subscript $_r$ for ``ring",
with $\xi_r=(2\xi_0^2-1)\eta_d - 2\xi_0\sqrt{\xi_0^2-1}\sqrt{\eta_d^2-1} $ and $ q_r=-\text{abs}\{((\xi_r^2-1)/(\eta_d^2-1))^{1/4}\}$, which are similar to the definitions of $\xi_h$ and $q_h$, but with $\xi_d\leftrightarrow\eta_d$.
 Substituting the approximation \eqref{Vr0b app} into \eqref{Vr inside} gives an approximate image potential $\mathring{V}_r^0$:
\begin{align}
	\mathring{V}_r^0&=2q_r\sum_{n=0}^\infty\sum_{m=-n}^n \frac{(n-m)!^2}{(n+m)!^2}(2n+1)~\Re\{ i^{m-1}Q_n^m(\xi_r)\}P_n^m(\xi_d)~ P_n^m(\xi)P_n^m(\eta)\cos m\phi \nonumber\\
	&=\text{sign}\bigg(\frac{z-\Re\{z_r\} }{\Im\{x_r\}}- \frac{x+\Re\{x_r\}}{\Im\{z_r\}}\bigg)\Re\bigg\{ \frac{-2i fq_r }{\sqrt{(x-x_r)^2+y^2+(z-z_r)^2}}\bigg\} \label{Vr0b inside}
\end{align}
where $x_r=-f\sqrt{\xi_r^2-1}\sqrt{\xi_d^2-1},~z_r=f\xi_r\xi_d$. 
Notice $\xi_d$ behaves as the $\eta$ coordinate of the image charge, and $\xi_d>1$ outside the usual range of $\eta$, while $\xi_r$ is complex. This results in $x_r$ and $z_r$ both being complex. The projection of the image onto real space is a ring similar to \eqref{V0center} but offset in the $x$ and $z$ directions, and tilted, such that the ring lies on the hyperbolic surface $\eta=\eta_d$, and at closest approach to the spheroid surface, the ring passes through the real image point charge \eqref{Vr0 inside}. The sign function makes the potential continuous through the hole in the ring. When the source is located at the center, \eqref{Vr0b inside} coincides with \eqref{V0center}. \\
We can then express the exact potential $V_r$ as a faster converging series by adding and subtracting the different forms of $\mathring V_r^0$:
\begin{align}
	V_r=&\text{sign}\bigg(\frac{z-\Re\{z_r\} }{\Im\{z_r\}}- \frac{x+\Re\{x_r\}}{\Im\{x_r\}}\bigg)\Re\bigg\{ \frac{-2i fq_r}{\sqrt{(x-x_r)^2+y^2+(z-z_r)^2}}\bigg\} \nonumber\\
	&-\sum_{n=0}^\infty\sum_{m=-n}^n \frac{(n-m)!^2}{(n+m)!^2}(2n+1) \bigg((-)^m\frac{Q_n^m(\xi_0)}{P_n^m(\xi_0)}P_n^m(\eta_d)+2q_r\Re\{i^{m-1}Q_n^m(\xi_r)\}\bigg)\nonumber\\
	&\hspace{7cm}\times P_n^m(\xi_d) P_n^m(\xi)P_n^m(\eta)\cos m\phi. \label{Vrbp in}
\end{align}

The approximations \eqref{Vr0 in lim} and \eqref{Vr0b app} apply only for $n\gg|m|$, although $m$ ranges from $-\infty$ to $\infty$. 
%But they are still useful because the terms decrease strongly in magnitude with increasing $|m|$, so are relatively insignificant. 
One way to quantify the rate of convergence of the double series numerically is to sum over $m$ first and then treat it as a single series over $n$. In this respect, \eqref{Vrp in} and \eqref{Vrbp in} improve the rate of convergence of the series a similar amount to that of the series \eqref{V'} for the point charge outside the spheroid, i.e. the terms converge faster by a factor of $n$.  The approximations are shown in Figure \ref{fig image real complex} for a point charge near the surface and near the axis, and roughly satisfy the boundary condition $V=0$ on the spheroid surface (their accuracy can be improved by a correction of the constant $n=0$ term). In terms of determining the analytic continuation of $V_r$, both $\dot V_r^0$ and $\mathring V_r^0$ exhibit a singular point at $\xi=\xi_h,\eta=\eta_d,\phi=0$, and this singularity is also partially visible in plots of the exact potential $V_r$ (partially due to the boundary of the domain of convergence $\xi=\xi_h$). 
%The ring image approximation $\mathring V_r^0$ suggests that the singularity of $V_r$ doesn't extend any further inwards from (decreasing $\xi$) from the ring, but this is only an approximation, and technically as far as we know the exact image singularity may lie anywhere where $\xi\geq\xi_h$. 
$\mathring V_r^0$ is also singular on the exterior disk that lies co-planar with the ring, but the exact analytic continuation of $V_r$ is not necessarily singular there.  %When the source charge is on the axis, $V_r$ shows a negative image ring on $\eta=\eta_d, \xi=\xi_h$ just as does the approximation $\mathring V_r^0$.

We can even combine the point image and ring image approximations to increase their accuracy. 
%Inspection of the series coefficients reveals that \eqref{Vr0 in lim} overshoots the coefficients while \eqref{Vr0b app} undershoots. 
By adding the two and weighting them according to how close the source is to the surface, the coefficients can be much more closely matched, and we can create a universal approximation for any source position (except the focal points -- see section \ref{sec focus}). A simple weighting may be 
\begin{align}
	\dot{\mathring{ V}}_r^0=\sqrt{\frac{\xi_d^2-1}{\xi_0^2-1}}\dot V_r^0 + \left(1-\sqrt{\frac{\xi_d^2-1}{\xi_0^2-1}}\right)\mathring V_r^0. \label{Vr0bb inside}
\end{align}
%where $\dot V_r^0$ is given in \eqref{Vr0 inside} and $\mathring{V}_r^0$ in \eqref{Vr0b inside}. 
The weighting factor $\sqrt{\xi_d^2-1}/\sqrt{\xi_0^2-1}$ is chosen so that near the surface $\xi_d\rightarrow\xi_0$ leaving $\dot{\mathring{  V}}_r^0\rightarrow \dot V_r^0$, while near the focus $\xi_d\rightarrow1$ leaving $\dot{\mathring{ V }}_r^0\rightarrow\mathring V_r^0$. This approximation is more accurate than $\dot V_0^r$ or $\mathring V_0^r$ separately, but there may be a more optimal weighting.

\subsection{Point charge on axis near the tip}
If the source charge is located on the $z$-axis with $z>f$, we have $\eta_d=1$. Here 
the ring approximation $\mathring V_r^0$ fails and the point charge approximation $\dot V_r^0$ reduces to a point charge on the axis.
%the approximation \eqref{Vr0 inside} reduces to 
%\begin{align}
%	\dot V_r^0&=q_h\sum_{n=0}^\infty(2n+1) Q_n(\xi_h)P_n(\xi)P_n(\eta) =\frac{fq_h}{\sqrt{\rho^2+(z-h)^2}}   \label{Vr0 tip}
%\end{align}
%which is a point charge on the $z$-axis. 
The series solution with improved rate of convergence \eqref{Vrp off axis} reduces to
\begin{align}
	V_r=&\frac{fq_h}{\sqrt{\rho^2+(z-h)^2}}
	-\sum_{n=0}^\infty(2n+1)\bigg(\frac{Q_n(\xi_0)}{P_n(\xi_0)}P_n(\xi_d) + q_hQ_n(\xi_h)\bigg) P_n(\xi)P_n(\eta). \label{Vrp tip}
\end{align}
where $h=f\xi_h$. We can attempt to analytically continue this by assuming that $V_r$ is singular on the $z$ axis for $z\geq h$. 
We follow a similar analysis to that done for the point charge on the axis outside the spheroid or in the center, that is, expand the potential on a basis of harmonic functions that share the same singularity as the reflected potential. In this case we use ``radially inverted irregular offset prolate spheroidal harmonics", or RIIOPSHs, which are also singular on the $z$-axis from $z=h$ to infinity.
Skipping the details, we can expand the series \eqref{Vrp in} in terms of regular spherical harmonics centered at the origin using the relationships between spheroidal and spherical harmonics (equation (2.21) of \cite{majic2016thesis}), then define a radially inverted radius $\hat r=h^2/r$ and treat this as the new radius, and expand these spherical harmonics onto offset irregular spheroidal harmonics (Eq. 2.28 of \cite{majic2016thesis}). The result is
\begin{align}
	V_r=\frac{fq_h}{\sqrt{\rho^2+(z-h)^2}}  -\sum_{k=0}^\infty (2k+1)&\sum_{p=0}^k \frac{(k+p)!}{p!^2(k-p)!} \sum_{n=p:2}^\infty (-)^{(n+p)/2+k} \frac{(n+p-1)!!}{p!(n-p)!!}\left(\frac{h}{f}\right)^{p+1}\nonumber\\
&\times (2n+1)\left( \frac{Q_n(\xi_0)}{P_n(\xi_0)}P_n(\xi_d)+q_hQ_n(\xi_h) \right)~ \frac{f}{r} Q_k(\hat{\xi})P_k(\hat\eta)  \label{Vr tip cont}
\end{align}
\begin{figure}
	\centering{\includegraphics[scale=.7]{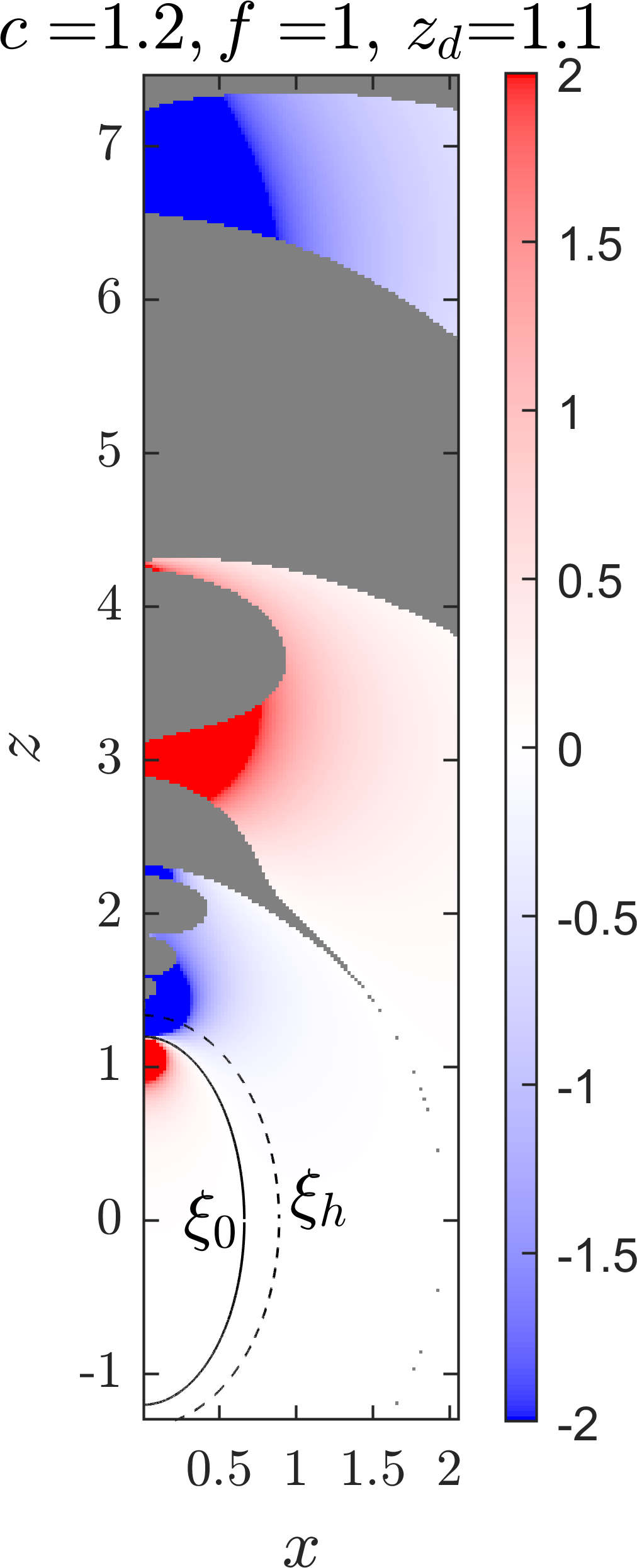}
		\includegraphics[scale=.7]{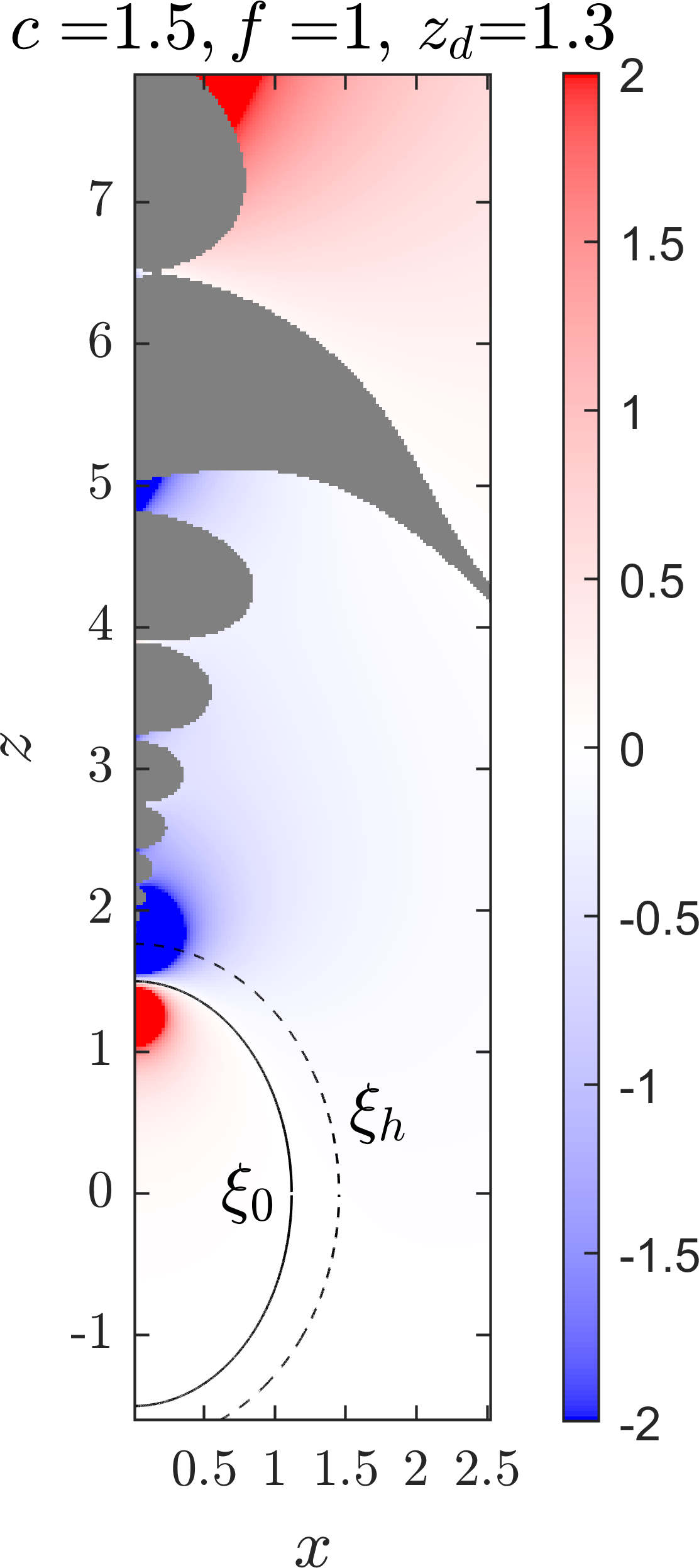}
		\includegraphics[scale=.7]{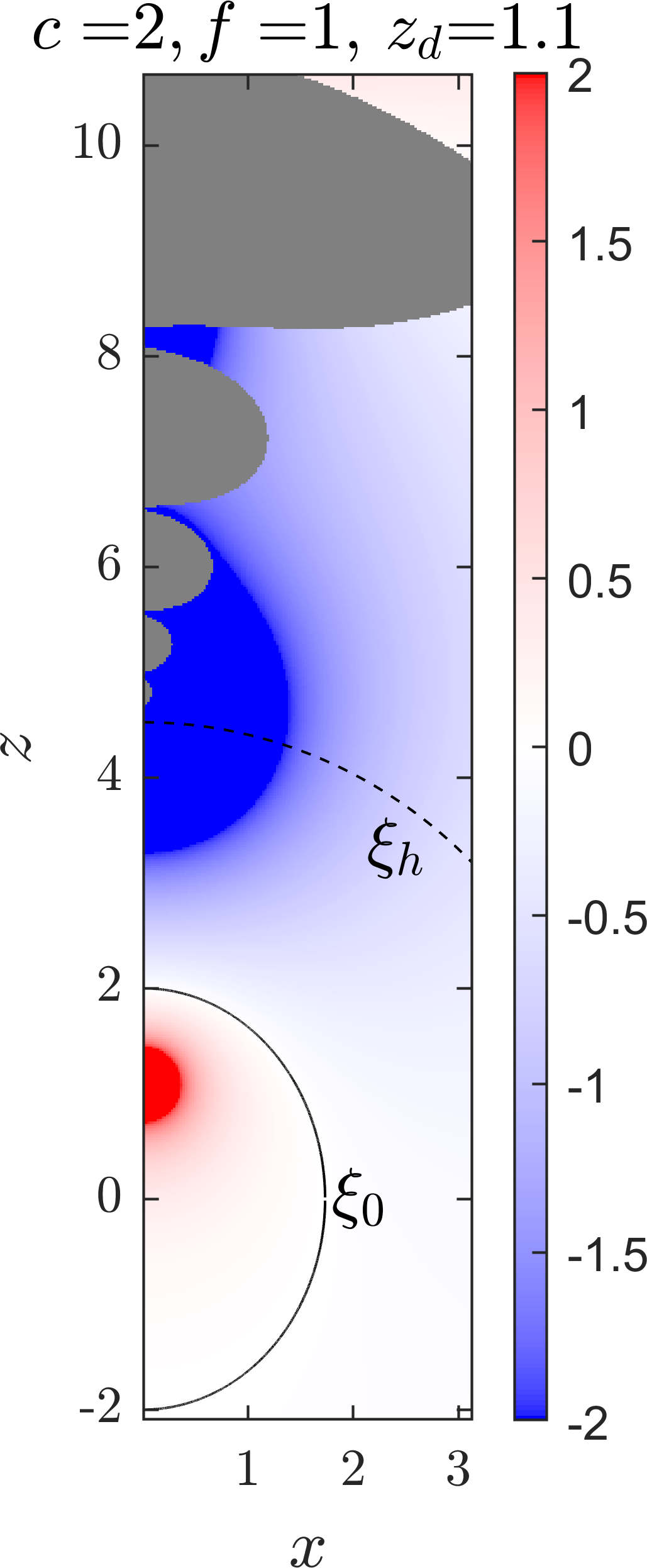}}\\
	\caption{Analytic continuation of the potential of a point charge inside a conducting spheroid, on the axis above the focus, for different aspect ratios and source positions. The potential was evaluated using the RIIOPSH series \eqref{Vr tip cont} with $n$ up to $n=100$ (enough for the sum over $n$ to converge). The grayed out area is where the potential has not converged sufficiently - the series was computed up to $k\leq K$ terms for two similar values of $K$; if these two series do not agree to within 10$\%$ at some point, that point is grayed out. For the left and right plots the $K$ values are $K=9,12$, and for the center plot $K=13,16$. } \label{fig Vr tip}
\end{figure}

where $\hat\xi=\xi(r\rightarrow \hat r),~\hat \eta=\eta(r\rightarrow\hat r)$. This series provides at least a significant analytic continuation of the potential $V_r$, as shown in Figure \ref{fig Vr tip}. The boundary of the domain of convergence of the RIIOPSH series is a coordinate of constant $\hat \xi$, which is a dimpled sphere where the dimple cuts into the sphere on the $z$-axis for $z>h$. Like the series \eqref{V''}, the series \eqref{Vr tip cont} suffers from numerical problems which are apparent near the singularity, prohibiting us from seeing the image charge distribution in detail. But the RIIOPSH series still provides a significant analytic continuation of the potential, and shows us where the singularities $aren't$ -- the continuation \eqref{Vr tip cont} can also be evaluated for large negative $z$ values and large $x$ values, and plots including these regions suggest that the singularity is concentrated to the $z$-axis for $z>h$. From the analogous problem for the ellipse (see Fig. \ref{fig ellipseimageinsidetiptip}), we could also guess that image contains singular points on the $z$-axis at $z=f\cosh(2k\acosh\xi_0\pm\xi_d)$ for $k=1,2,3...$.

\FloatBarrier
\subsection{Point charge on focal point}\label{sec focus}
\begin{figure}
	\includegraphics[scale=.8,trim=0 -.14cm 0 0 , clip]{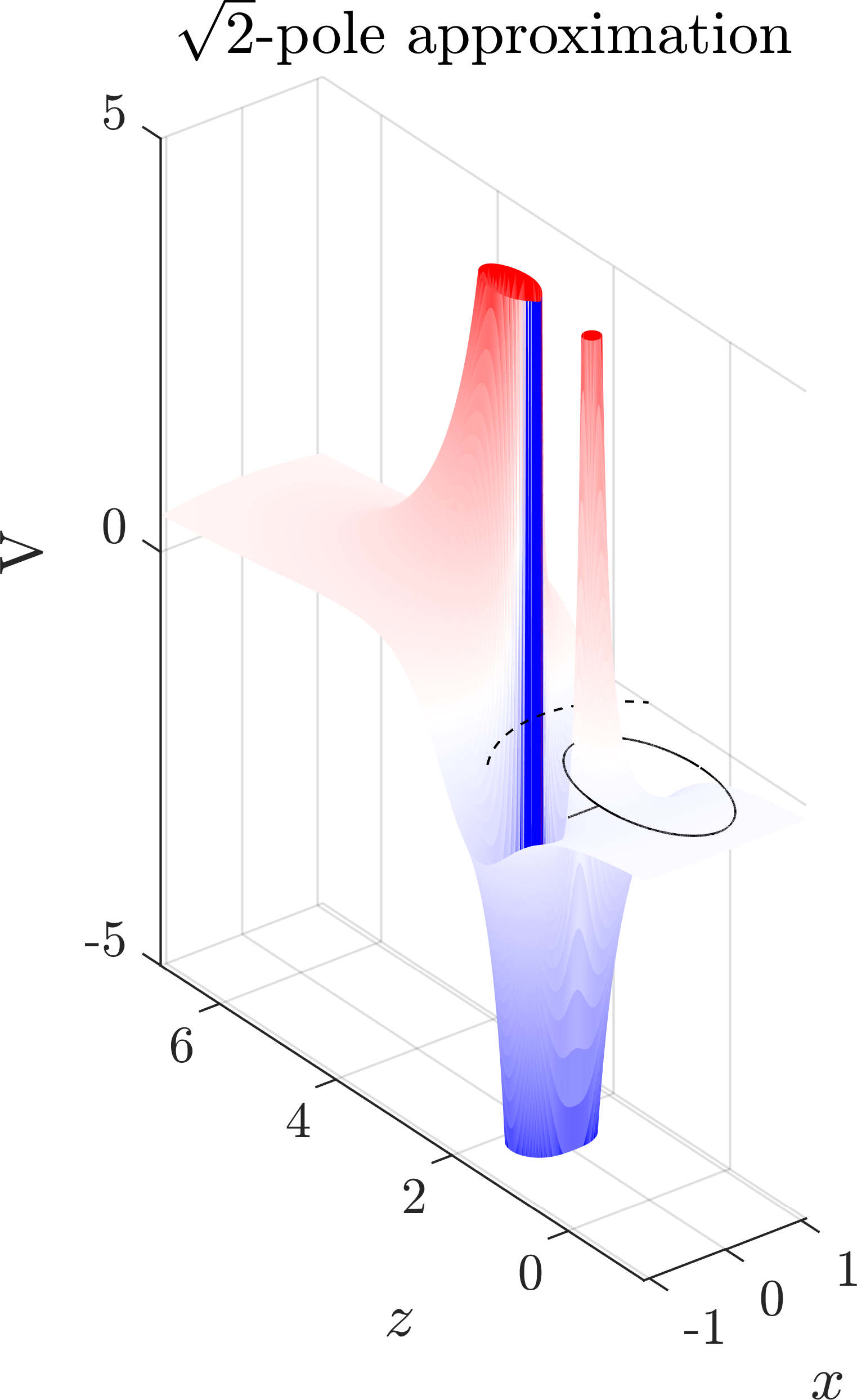}
		\includegraphics[scale=.8,trim=0 -.14cm 0 0 , clip]{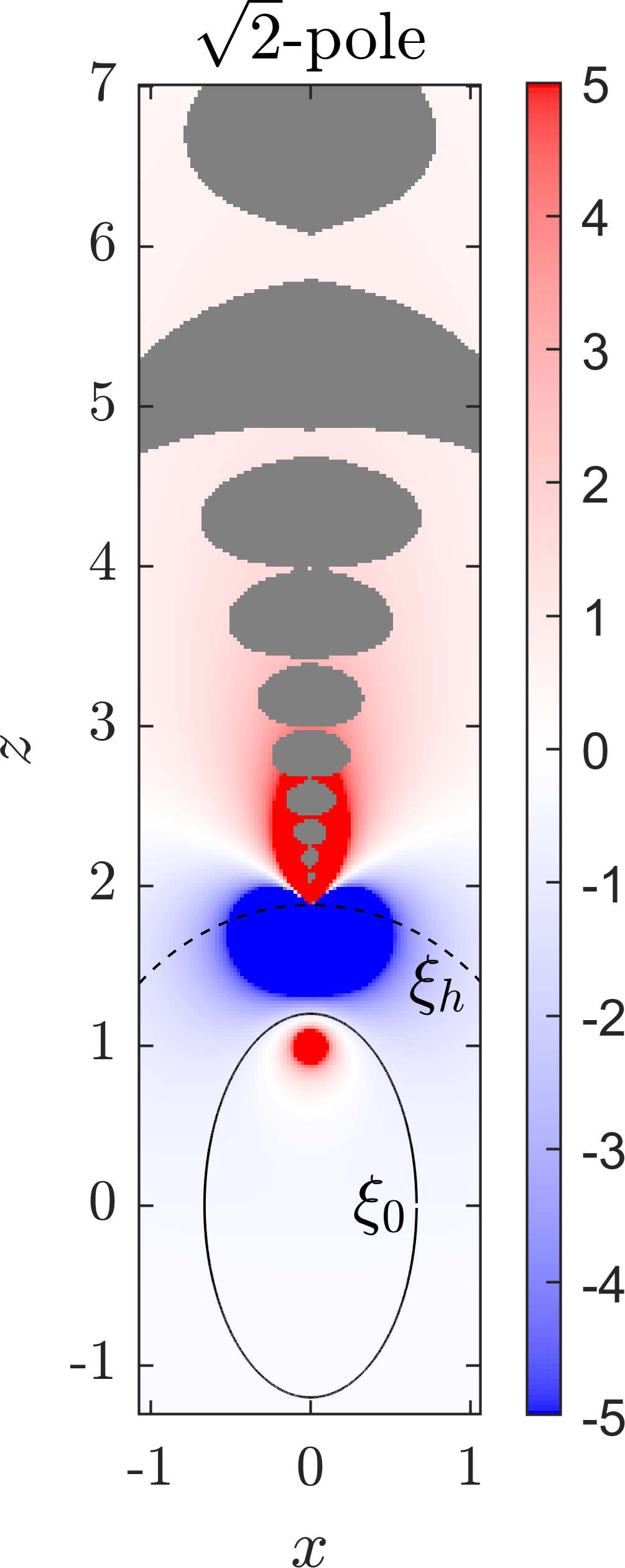}
\includegraphics[scale=.8,trim=-.5cm 0cm 0 0 , clip]{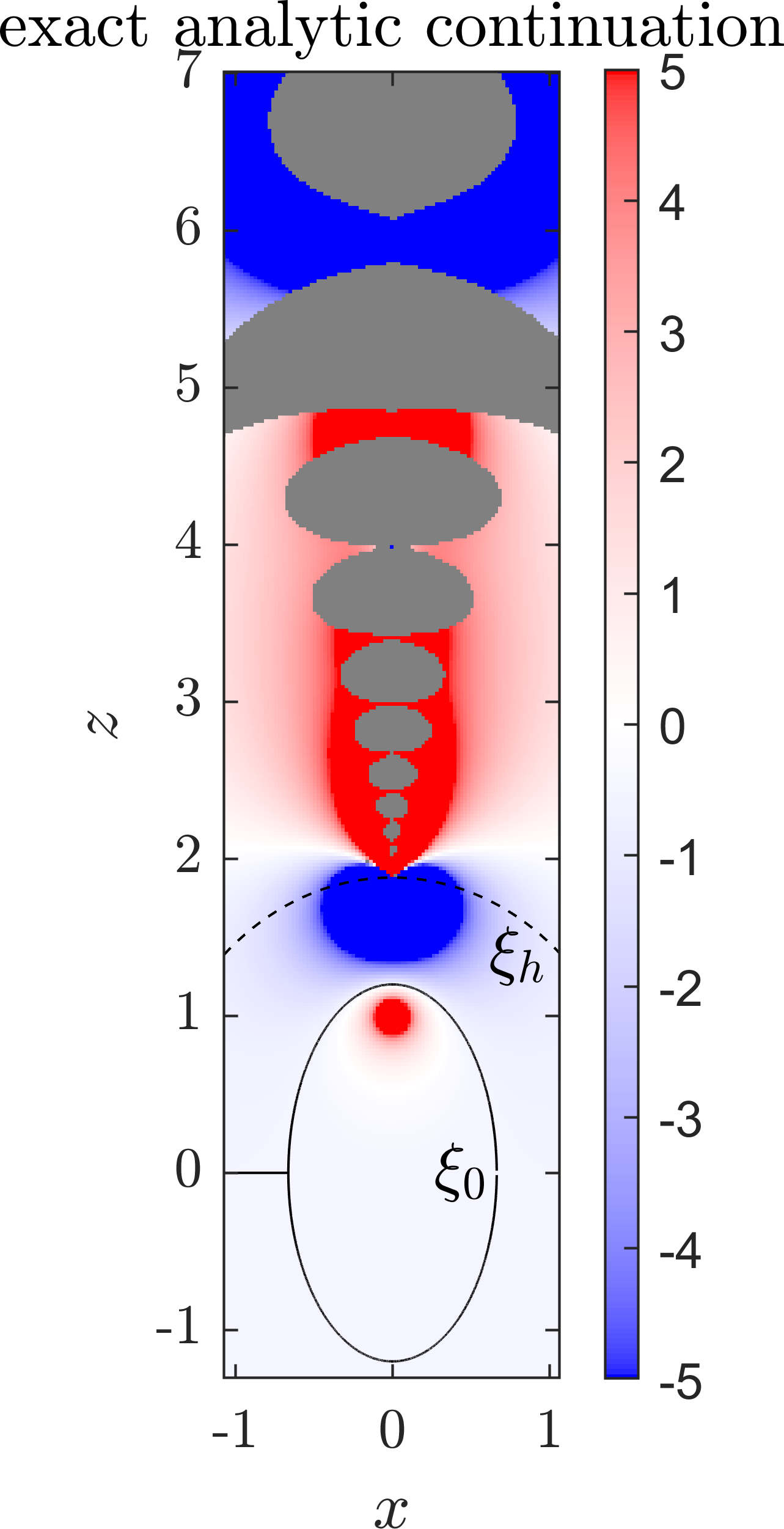}
		\caption{Left: Height map of the approximation $V_e+V_{\sqrt{2}}^h$ for a point charge located on the top focal point of the conducting spheroid of half-height $c=1.2$, half-focal length $f=$1.
Right: color plot of the analytic continuation of the exact potential for the same problem, evaluated using the RIIOPSH series \eqref{Vr focus cont} up to $n=100$ (enough for the sum over $n$ to converge) and $k=20$. The grayed out area is where the potential has not converged sufficiently - the series was compared to itself computed with only $k=17$ terms; an area is grayed out if these two series do not agree to within 10$\%$.} \label{fig Vf}
\end{figure}

For a source charge on the top focal point, i.e. $\xi_d=1,~\eta_d=1,~z_d=f$, both approximations \eqref{Vr0 inside} and \eqref{Vr0b inside} fail due to the factor $(\xi_d^2-1)^{-1/4}$ in $q_h$ and $(\eta_d^2-1)^{-1/4}$ in $q_r$. Here the spheroidal harmonic solution is 
\begin{align}
	V_r^f=-\sum_{n=0}^\infty (2n+1)\frac{Q_n(\xi_0)}{P_n(\xi_0)}P_n(\xi)P_n(\xi). \label{Vr focus}
\end{align}
The coefficients for $n\rightarrow\infty$ go as
\begin{align}
	\frac{Q_n(\xi_0)}{P_n(\xi_0)}\rightarrow \pi \bigg(\xi_0+\sqrt{\xi_0^2-1}\bigg)^{-2n-1} \big[1+\cO(1/n)\big], \label{lim focus}
\end{align}
but unlike in the previous sections we cannot easily match this behavior simply with one or two Legendre functions in the numerator because for $n\rightarrow\infty$, both $P_n$ and $Q_n$ have a factor of $n^{-1/2}$ which is not present in \eqref{lim focus}. Instead we find that a good approximation involves the fractional order Legendre function $Q_n^{1/2}$ which does not have this $n^{-1/2}$ dependence. The approximate potential is that of a fractional multipole of degree 1/2, or a ``$\sqrt{2}$-pole", which in a sense is half way between a monopole and a dipole. The theory of fractional multipoles and how to create them using fractional derivatives is investigated in \cite{engheta1996fractional}. They define in their equation (30) the multipole of degree $\alpha$ by applying the fractional derivative $_{-\infty}\! D_z^\alpha$ to a monopole at the origin (where $-\infty$ is a lower value like that used in integration, $z$ is the variable to be differentiated, and $\alpha$ is the order of differentiation). For $\alpha=1/2$, with our normalization, this is
\begin{align}
	V_{\sqrt{2}}&=_{-\infty}\!D^{1/2}_z \frac{\ell^{1/2}}{\sqrt{\rho^2+z^2}} \label{V1/2 D}\\
    &=\frac{\ell^{1/2}}{\Gamma(1/2)}\frac{\pd}{\pd z}\int_{0}^\infty \frac{v^{-1/2}}{\sqrt{\rho^2+(z-v)^2}}\d v \label{V1/2 int}\\
	&= \frac{\ell^{1/2}\Gamma(3/2)}{r^{3/2}}P_{1/2}(-\cos\theta). \label{V1/2}
\end{align}
where $\ell$ is an arbitrary length. $V_{\sqrt{2}}$ is a solution to Laplace's equation except on the positive $z$-axis due to the Legendre function $P_{1/2}$ being singular at $\cos\theta=1$. The line charge density of this multipole is $-z^{-3/2}$ for $z>0$, plus a positive point charge at the origin. 
What we want for an approximate image is $V_{\sqrt{2}}$ but translated up the $z$-axis, so that the singularity extends upwards from the height where the series \eqref{Vr focus} diverges, that is $\rho=0,z=h$, where $h=\xi_h f$ and $\xi_h=2\xi_0^2-1$. We shall call this approximation $V_{\sqrt{2}}^h$. It turns out that the coefficients of the spheroidal harmonic expansion of this image have exactly the behavior of the coefficients in \eqref{lim focus}. 
In order to derive these coefficients we begin with the known expansion of the monopole: 
\begin{align}
	V_1^h=
	\frac{f}{\sqrt{\rho^2+(z-h)^2}}=\sum_{n=0}^\infty (2n+1) Q_n(\xi_h)P_n(\xi)P_n(\eta), \label{V0mono focus}
\end{align}
and note that the symmetry about $z\leftrightarrow h$ means that we can derive the $\sqrt{2}$-pole by differentiating the monopole with respect to $h$ instead of $z$. Applying this to the expansion \eqref{V0mono focus} and modifying the prefactors so that the series coefficients match those in \eqref{lim focus} leads to
\begin{align}
%	V_\sqrt{2}^h=&\frac{(\xi_h^2-1)^{3/4}}{2}\!D_h^{1/2}\frac{f}{\sqrt{\rho^2+(z-h)^2}} ***check***\nonumber\\
	V_{\sqrt{2}}^h=&-\sqrt{2\pi f(\xi_h^2-1)} ~~_{-\infty}\!D_h^{1/2} V_1^h\nonumber\\
           =&-\sqrt{2\pi f(\xi_h^2-1)}\sum_{n=0}^\infty (2n+1) ~_{-\infty}\!D_h^{1/2}\big[Q_n(\xi_h)\big] P_n(\xi)P_n(\eta).
\end{align}
The fractional derivative of $Q_n(\xi_h)$ can be evaluated using formula 7.133 from \cite{tables2014}, %pg. 780
 which gives the result in terms of a Legendre function of order $m=1/2$:
\begin{align}
	V_{\sqrt{2}}^h=-\sqrt{2\pi}(\xi_h^2-1)^{1/4}\sum_{n=0}^\infty (2n+1) Q_n^{1/2}(\xi_h)P_n(\xi)P_n(\eta), \label{V0 focus}
\end{align}
Conveniently, the Legendre function $Q_n^{1/2}$ has a simple formula:
\begin{align}
	Q_n^{1/2}(\xi_h)=\frac{\sqrt{\pi/2}}{(\xi_h^2-1)^{1/4}} \left(\xi_h+\sqrt{\xi_h^2-1}\right)^{-n-1/2}.
\end{align}
The closed form expression of $V_{\sqrt{2}}$ is  
\begin{align}
	V_{\sqrt{2}}^h=-\frac{\pi}{\sqrt{2}}\frac{\sqrt{\xi_h^2-1}f^{3/2}}{(\rho^2+(z-h)^2)^{3/2}} P_{1/2}(-\cos\theta_h), \label{V12h}
\end{align}
where $\cos\theta_h=(z-h)/\sqrt{\rho^2+(z-h)^2}$, and the Legendre function $P_{1/2}$ can be evaluated via its hyper-geometric series. The approximate solution $V_e+V_{\sqrt{2}}^h$ is plotted in Fig. \ref{fig Vf}, showing a negative image point charge at $z=h$ and a positive line charge distribution for $z>h$, decaying as $z$ increases. 
How similar is the singularity of the exact solution?
%Extracting \eqref{V0 focus} from the series \eqref{Vr focus} makes the series coefficients converge faster by a factor $1/n$:
%\begin{align}
%	V_r^f=	V_{\sqrt{2}}^h -\sum_{n=0}^\infty (2n+1)\left( \frac{Q_n(\xi_0)}{P_n(\xi_0)}+\sqrt{2\pi}(\xi_h^2-1)^{1/4}\right) Q_n^{1/2}(\xi_h)P_n(\xi)P_n(\eta). \label{Vrp focus}
%\end{align}
We can guess that the potential is singular on the $z$ axis for $z\geq h$, and obtain its analytic continuation following a similar analysis to that done for the point charge on the axis for $z>f$, but starting with the series \eqref{Vr focus} (also to minimize numerical cancellations we add \eqref{V12h} and subtract \eqref{V0 focus}). The result is
\begin{align}
	V_r^f=V_{\sqrt{2}}^h -\sum_{k=0}^\infty (2k+1)&\sum_{p=0}^k \frac{(k+p)!}{p!^2(k-p)!} \sum_{n=p:2}^\infty (-)^{(n+p)/2+k} \frac{(n+p-1)!!}{p!(n-p)!!}\left(\frac{h}{f}\right)^{p+1}\nonumber\\
	&\times (2n+1)\left( \frac{Q_n(\xi_0)}{P_n(\xi_0)}-\sqrt{2\pi}(\xi_h^2-1)^{1/4}Q_n^{1/2}(\xi_h)\right) \frac{f}{r} Q_k(\hat{\xi})P_k(\hat\eta). \label{Vr focus cont}
\end{align}
Like series \eqref{Vr tip cont}, this suffers from numerical problems for $k\gtrsim20$, and is plotted in Figure \ref{fig Vf} (right) with as much accuracy as possible. It appears that the approximation $V_{\sqrt{2}}^h$ closely matches the singularity of $V_r$ near $z\approx h$.
The main visible difference between the images of a point charge on the focal point $z_d=f$ or above the focus $f<z_d<c$, is that for $z_d=f$ the image line charge density is positive for $z\gtrsim h$, but for $z_d>f$ the image line charge is negative for $z\gtrsim h$, then becomes positive higher up the axis.
From the analogous problem for the ellipse, we could also guess that image contains singular points on the $z$-axis at $z=f\cosh(2k\acosh\xi_0)$ for $k=1,2,3...$.
It is interesting to note that the innermost point of the image at $z=h$ is the same as the critical distance $d_c$ (Eq. \eqref{dc}) for the source charge located outside the spheroid on the axis. For a source at $z_d=d_c$, the approximate point image \eqref{V^(0)} lies exactly on the focal point. 
%So the interior and exterior problems for point charges located at $z=f$ or $z=h=d_c$ are in a sense inverses of each other, at least approximately in terms of where their leading order image point charges lie. 

\section{Conclusion} \label{sec conclusion}

We have investigated obtaining reduced images of point sources inside or outside ellipses and prolate spheroids, manipulating the series solutions to reveal the singularities of the reflected potential, and deriving approximate images that mimic these singularities. 
For the ellipse, simple reduced images could be found for every position of the source. For an exterior source the problem was solved via series \cite{sten1996focal} and reflection \cite{alshal2021image}, and both could provide analytic continuation of the potential to reveal the image on the focal line, plus a point image if the source was close. For an internal source, we derived a solution as a series of image charges lying on a hyperbola.

For a point charge on the axis of a prolate spheroid, within the critical distance to the surface, numerical analysis of a series of stretched spheroidal harmonics \eqref{V''} suggests that the reduced image lies on the segment extending from the lower focus up to the image point: $-f<z<h$, and that the top point of the singularity is well matched by the point charge approximation of \cite{lindell2001electrostatic}. For a point charge off the axis, we have generalized this point charge approximation with the requirement that the coefficients of its series expression have the same asymptotic form as those for the reflected potential. 

For a point charge inside the spheroid, we find various image approximations. 
For a point charge at the center of a prolate spheroid, we also analytically continued the potential using a series of radially inverted oblate spheroidal harmonics, which suggests that the exact image lies on the external disk $z=0$, $\rho\geq h$. A similar analysis for the source charge on the axis above the focus was carried out using radially inverted offset prolate spheroidal harmonics which suggests the exact image lies on the $z$-axis for $z>h$.
For the source charge near the surface, an image point charge opposite the surface is  approximately valid. For the source charge near the focal segment, a point charge in complex space or equivalently an image ring/exterior disk is appropriate. For the source exactly on one of the focal points, the image approximation is found to be a $\sqrt{2}$-pole consisting of a line charge extending up the $z$-axis.

%Unfortunately all the series used here to reveal the exact analytic continuations have severe numerical problems. Future work to improve the numerical stability of these series or evaluate them with higher precision would allow us to visualize the image in more detail. 

%We also cannot analytically prove that the series actually converges everywhere except this disk. 

%We have considered the concept of uniqueness of image solutions, and noted that surface images are in a sense non-unique. First of all, any closed surface can be analytically continued through to reveal a more reduced image. But even an image lying on an open surface is not unique. We presented two functions that solve problem of a grounded oblate spheroid in a uniform potential, whose corresponding image surface charges are different.

%A future goal is to find the form of the image for an off-axis point source. An analytic approach could be to use a Watson transformed series of the type \eqref{V nu}.
\begin{acknowledgments}
	I wish to thank Victoria University for funding this research via a doctoral scholarship, and an anonymous reviewer for many helpful suggestions.\\
\end{acknowledgments}

%\textbf{Data availability.}
%The data that support this study are available upon request.

\thispagestyle{empty}
\bibliographystyle{ieeetr}
%\bibliography{libraryImageSpheroid}
\bibliography{../libraryH}
\end{document}